%% file: GaussReversals.tex
\documentclass{elsart}
\usepackage{graphicx}
\usepackage{amsmath}
\usepackage{natbib}
\usepackage{lscape}

\include{abb}

\begin{document}

\begin{frontmatter}

\title{A Gaussian Model for Simulated Geomagnetic Field Reversals}
\author{Johannes Wicht$^1$ and Domenico Meduri$^{1,2}$}
\address{$^1$Max-Planck-Institut f\"ur Sonnensystemforschung, G\"ottingen, Germany}
\address{$^2$University G\"ottingen, Germany}
\ead{Wicht@mps.mpg.de}

\begin{abstract}
Field reversals are the most spectacular changes in the geomagnetic field
but remain little understood.
Paleomagnetic data primarily constrain the reversal rate and provide
few additional clues. Reversals and excursions are characterized
by a low in dipole moment that can last for some $10\,$kyr.
Some paleomagnetic records also suggest that the field decreases
much slower before an reversals than it recovers afterwards and that
the recovery phase may show an overshoot in field intensity.
Here we study the dipole moment variations in several
extremely long dynamo simulation to statistically
explored the reversal and excursion properties.
The numerical reversals are characterized by a switch from
a high axial dipole moment state to a low axial dipole moment state.
When analysing the respective transitions we
find that decay and growth have very similar time scales and that there is no overshoot.
Other properties are generally similar to paleomagnetic findings.
The dipole moment has to decrease to about $30$\% of its mean to
allow for reversals. Grand excursions during which the field intensity
drops by a comparable margin are very similar to reversals and likely have the
same internal origin. The simulations suggest that both are simply
triggered by particularly large axial dipole fluctuations while other
field components remain largely unaffected.
A model at a particularly large Ekman number shows a second but little
Earth-like type of reversals where the total field decays and recovers
after some time.
\end{abstract}


\end{frontmatter}

\input{Introduction2}
\input{Methods2}

\input{DynamoModels2}
\clearpage
\input{Discussion}


\input{GaussReversals.bbl}

\end{document}

%% file: abb.tex
\newcommand{\refp}[1]{(\ref{#1})}

\newcommand{\B}{\mbox{$\mathbf B$}}
\newcommand{\U}{\mbox{$\mathbf U$}}

\newcommand{\be}{\begin{equation}}
\newcommand{\ee}{\end{equation}}
\newcommand{\eep}{\;\;.\end{equation}}
\newcommand{\eec}{\;\;,\end{equation}}
\newcommand{\bea}{\begin{eqnarray}}
\newcommand{\eea}{\end{eqnarray}}
\newcommand{\bel}[1]{\begin{equation}\label{#1}}
\newcommand{\beal}[1]{\begin{eqnarray}\label{#1}}
\newcommand{\tp}[1]{\mbox{$\times10^{#1}$}}

\newcommand{\Ro}{\mbox{Ro}}

\newcommand{\E}{\mbox{E}}

\newcommand{\Pra}{\mbox{Pr}}
\newcommand{\Pm}{\mbox{Pm}}
\newcommand{\Rm}{\mbox{Rm}}

\newcommand{\Ra}{\mbox{Ra}}

\newcommand{\Els}{\mbox{$\Lambda$}}

\newcommand{\Dgeo}{\mbox{D$_{\ell\le 11}$}}

\newcommand{\TC}{{\tau}}
\newcommand{\dur}{\TC_V}
\newcommand{\wai}{\TC_W}
\newcommand{\lev}{\TC_L}
\newcommand{\dec}{\TC_D}
\newcommand{\gro}{\TC_G}
\newcommand{\trans}{\TC_T}
\newcommand{\all}{{\TC_A}}
\newcommand{\tdur}{t_V}
\newcommand{\twai}{t_W}
\newcommand{\tlev}{t_L}
\newcommand{\tdec}{t_D}
\newcommand{\tgro}{t_G}
\newcommand{\tdip}{t_d}

\newcommand{\tsv}{t}
\newcommand{\MC}{{\mathcal M}}
\newcommand{\MCH}{\MC_{Hc}}
\newcommand{\MCL}{\MC_{Lc}}
\newcommand{\MH}{M_{Hc}}
\newcommand{\ML}{M_{Lc}}
\newcommand{\MM}{\ovl{M}}

\newcommand{\ElsCMB}{\Lambda_{\mbox{{\scriptsize CMB}}}}

\newcommand{\ovl}[1]{\overline{#1}}
\newcommand{\corr}{\mathcal{C}}

\newcommand{\Figref}[1]{Fig.~(\ref{#1})}
\newcommand{\figref}[1]{fig.~(\ref{#1})}

\newcommand{\eqnref}[1]{eqn.~(\ref{#1})}
\newcommand{\Secref}[1]{Section \ref{#1}}
\newcommand{\secref}[1]{section \ref{#1}}

\newcommand{\tabref}[1]{table {\ref{#1}}}
\newcommand{\Tabref}[1]{Table {\ref{#1}}}

%% file: Introduction2.tex
\section{Introduction}
\label{Intro}
The geomagnetic field varies over a broad range of time scales
raging from one to many million years \citep{Constable2005}.
Adequately covering these vastly different time scales is a
challenge for geomagnetic data acquisition and numerical
dynamo modelling alike.
Here we mostly focus on variations in the dipole component
and in particular on reversals and excursions.
Analysis of Earth's secular variation in historical and archeomagnetic field
models suggests that the typical dipole time scale is about a millennium
while higher field harmonics change over centuries or decades
\citep{Christensen2004,Lhuillier2011}.
Paleomagnetic records constrain the longer time scales but suffer from
a lack of temporal resolution, ambiguities in dating,
and difficulties in determining the paleofield intensity \citep{Hulot2010}.
Slow dipole variations are mostly associated to reversals and excursions,
a second class of events where the geomagnetic pole ventures further away
from the geographic pole but ultimately returns.
The associated changes in the dipole tilt
last from one to several millennia \citep{Clement2004,Valet2012}.
Both types of events are characterized by a strong decrease
in field intensity likely caused by a drop in axial dipole strength
which can last several $10\,$kyr \citep{Valet2005,Singer2005,Channell2009,Amit2010b,Valet2012}.
The last reversal, the transition from the Matuyama to the Brunhes polarity chron,
happened about $780\,$kyr ago but reversals are stochastic events and their rate has varied
greatly over the paleomagnetic history. A particularly long period
without any reversal is the Cretaceous normal superchron that
lasted about $35\,$Myr but up to $10$ reversals per Myr have also
been recorded. Variations in the reversals rate are typically associated to
changes in Earth's lower mantle which have a similar time scale of some $10\,$Myr \citep{Biggin2012}.

Excursions seem to be considerably more frequent than reversals.
Up to $17$ excursions have been reported for the Bruhnes \citep{Lund2001,Lund2006,Roberts2008}
and up to $9$ for the Matuyama chron \citep{Channell2009}.
Since reversals are bound by longer lasting chrons
with opposite polarity they are relatively easy to detect in a
paleomagnetic sequence.
The exact timing of the reversal may still be difficult but the presence of the
chrons is evidence enough. Excursions, however, are only recorded when
the temporal resolution of the paleomagnetic medium
exceeds their relatively short duration \citep{Roberts2004}.
The number of documented excursion will likely increase when better quality
high-resolution records become available \citep{Roberts2008}.
Additional uncertainties arise from the fact that some excursions may not be global events.

Paleomagnetic records provide local rather than global information.
While the local field remains a good proxy for the dipole behaviour
during stable polarity epoches this is not necessarily true during reversals and excursions
when the relative dipole contribution is low.
The likelihood of recording transitional or inverse directions at a given site therefore
depends on the multipole field contributions and also increases with duration and depth of the drop in dipole
intensity \citep{Wicht2005,Wicht2009}.
A preponderance of sites with inverse directions during
recent excursions could mean that the dipole had started to grow with
inverse polarity during these events \citep{Valet2008}.
Excursions may thus be abbreviated reversals.
\citet{Gubbins1993b} suggests that an inverse direction can only be established
when the inverse field has a chance to diffuse deep into the inner core.
Excursions are then simply events where the period of inverse field production is
shorter than the inner core dipole diffusion time of about $7\,$kyr.
Dynamo models by \citep{Wicht2002}, on the other hand, indicate that the highly
dynamic dynamo process cares little for the dissipation in Earth's relatively small
inner core but newer simulations addressing this question show contradictory
results \citep{Dharmaraj2012,Lhuillier2013}.


Reversals during the last $2\,$Myr are characterized by a drop in dipole moment
from a mean value of around $\ovl{M}=7\tp{22}$A$\,$m$^2$ to a low around
$M_L=2\tp{22}$A$\,$m$^2$ \citep{Valet2005,Channell2009}.
When analysing high quality sedimentary data for the five most recent reversals,
\citet{Valet2005} find that the magnetic field strength decays much slower before actual
polarity transition than it grows afterwards. The decay phase lasts about $50\,$kyr and
is thus comparable to the core dipole decay time of $\tdip\approx 56\,$kyr
based on newly revised estimates for the electrical conductivity of Earth's core alloy \cite{Pozzo2012}.
While the decrease in magnetic field strength could thus be a simple diffusive process this
is not the case for the faster recovery  which takes less than $10\,$kyr.
Some records also suggest that the field intensity overshoots its mean value at the end
of the recovery phase \citep{Petrelis2009}.
Neither a significantly faster recovery nor the overshoot
have ever been reported for excursions but the drop in field intensity during these events
seems comparable to those recorded for reversals \citep{Roberts2008,Channell2009}.
\citet{Ziegler2011a}, however, report that magnetic field decayed on average $20$\% faster
than it growed during the last $2\,$Myr when considering time scales longer than about $25\,$kyr.
The strong site dependence and complex transition behavior during
reversals and excursions indicates that the more time dependent higher harmonic contributions
dominate at least for some time \citet{Singer2005,Valet2005,Valet2012}.
The drop in field intensity is thus mainly caused by a decrease of the dipole component.


The cause for reversals and excursions remains little understood.
Insight comes from paleomagnetic measurements, theoretical considerations,
laboratory experiments, and numerical dynamo simulations.
Simple dynamical models that manage to reproduce their fundamental stochastic nature
include Rikitake coupled disc dynamos
\citep{Rikitake1958}, mean field dynamos \citep{Ryan2007,Stefani2007}, low order dynamical models
\citep{Melbourne2001,Schmitt2001,Petrelis2009}, or a coupled spin system \citep{Mori2013}.

The VKS dynamo is the only experiment that features Earth-like reversals and excursions,
including a faster recovery and an overshoot after reversals but not after excursions
\citep{Ravelet2008}.
The low order model by \citet{Petrelis2009} manages to explain these differences despite
reducing the dynamics to the interaction between only two magnetic field components.
The primary component is the axial dipole while the secondary
component can be any other field contribution.
Configurations with dominant primary field of either polarity
form two stable points in phase space that are supplemented by two
unstable points with weaker primary and stronger secondary field.
Random fluctuations can drive the system away from a stable point towards the
closest unstable point, causing a decrease in the axial dipole.
When the fluctuations team up constructively enough to cross the unstable point,
the system is attracted towards the opposite stable point and
a reversal happens. The stochastic variation around
the stable point can be understood as a random walk and are thus comparatively slow \citep{Schmitt2001}.
The attraction by the opposite attractor beyond the unstable points guarantees a faster recovery
and an overshoot towards the second unstable point.
Fluctuations short of the unstable points can lead to excursions where
the system returns to the original fix point.
Both the axial dipole decay and growth phase of an excursion would
be slow and there is not overshoot.
This dynamics is typical for a system just below a saddle node
bifurcation. Beyond the bifurcation, however, stable and unstable points coincide
and the system undergoes limit cycles equivalent to continuous reversing \citep{Petrelis2009}.

Caution is required when transferring the results from such simplified approaches
to the geodynamo. Full 3d dynamo simulations feature a more realistic dynamic but
suffer from limitations in computing power and problems in interpreting the
complex dynamics.
The Ekman number, a measure for the relative importance of viscous forces,
has to be chosen many orders of magnitude too high to damp the
smaller scales that cannot be resolved with the available computing power.
This is particularly true for reversal simulations that have to be
integrated sufficiently long enough to capture these rare events.
\citet{Takahashi2005} present a reversing model
at an  Ekman number of $\E=2\times 10^{-5}$. More typical, however,
are larger values up to $2\times 10^{-2}$ \citep{Wicht2005,Aubert2008a}.
Earth's Ekman number is around $10^{-15}$ (when base on molecular diffusivities)
while most advanced dynamo simulations reach $10^{-6}$ but are too short or too
weakly driven to show reversals \citep{Wicht2010b}.

Numerical parameter studies suggest that sufficiently strong inertial forces are essential
for triggering reversals \citep{Christensen2006,Driscoll2009,Wicht2009,Olson2014}.
\citet{Christensen2006} demonstrate that this can be quantified in
term of the  local Rossby number
\bel{Roll}
\Ro_\ell = \mbox{U} / (\Omega d)\;\;.
\ee
This modified Rossby number, based on the rms flow amplitude U and a characteristic flow length scale $d$,
is a measure for the ratio of inertial to Coriolis forces in the Navier-Stokes equation.
The dipole field clearly dominates and is very stable as long as reversal forces are weak.
Beyond about $\Ro_\ell\approx 0.1$, however, the dipole seems to loose its special role, becomes comparable to
the other multipole contribution, and reverses more or less continuously.
Earth-like reversals where the axial dipole still dominates in a mean sense and reversals and excursions
are rare can be found in a narrow $\Ro_\ell$ interval just before the transition \citep{Wicht2009}.
These events thus happen close to a critical point in the parameter space where the system changes its dynamics,
a concept discussed by several of the low order dynamical models \citep{Stefani2007,Petrelis2009}.
Fluctuations in the highly variable flow and magnetic field cause transitions from the stable into the multipolar
regime where the dipole reverses more easily \citep{Kutzner2002}.
When the system returns, it is just a matter of chance whether the original or the inverse polarity is amplified.
Reversals and excursion should thus be equally likely \citep{Wicht2009}.
In terms of the simple low order approach by \citet{Petrelis2009} these events would be described by
the limit cycle process flanked but additional dynamical processes that allow to cross the bifurcation.
Faster recovery and overshoot may still be possible but am inherent feature of such a model.

No full dynamo simulation has directly addressed the question of
a possible asymmetry in dipole decay and growth during reversals or excursions.
\citet{Lhuillier2013} analyse five reversing dynamos at an
Ekman number of $\E=6.5\times 10^{-3}$ with altogether nearly $2000$ reversals.
Judging from their figures, neither model shows any faster recovery but
one features at least the overshoot. Unfortunately, this model is little Earth-like since it
uses an electrically insulating inner core and shows much fewer excursions than reversals.
\citet{Gissinger2010} explore a simplified dynamo model where Coriolis force and
buoyancy force are replaced by a suitable driving term that produces a reversing dipole dominated field.
A faster decay can only be observed when the magnetic Prandtl number $\Pm$, the
ratio of viscous to magnetic diffusivity, is below unity.
Since the magnetic Prandtl number where dynamo action is still possible decreases with
the Ekman number in full dynamo models a $\Pm$ below unity can only be reached at $E\le 10^{-4}$
\citep{Christensen2006}.
The model by \citet{Takahashi2005} is the only one in this range and neither the
faster recovery not the overshoot are reported.
However, their numerical runs cover only few transitions and these issues are not directly addressed.

Most simulations show that the reversal behavior can be very complex and highly variable in time
\citep{Glatzmaier1999,Coe2000,Kutzner2002,Wicht2005,Wicht2009,Driscoll2009,Lhuillier2013}.
Addressing the properties of reversals and excursions therefore requires a statistical
approach based on particularly long simulations with as many simulations as
possible. The reversal behavior depends on the convective driving and the
thermal boundary conditions used in the numerical model, for example on the outer-boundary
heat flux pattern imposed by the lower mantle \citep{Glatzmaier1999,Kutzner2004,Olson2010,Olson2013}.
System parameters like Ekman number and Rayleigh number also play an important role
\citep{Driscoll2009,Olson2014}.

The main purpose of this paper is to analyse the statistical properties of reversals and excursions
in several dynamo models and explore whether they reproduce paleomagnetic observations.
The models cover Ekman numbers from $\E=2\tp{-2}$ to $\E=3\tp{-4}$ and also
differ in Rayleigh number and convective driving mode.
We start with explaining the numerical approach in \secref{Methods} and
introduce the different dynamo models in \secref{Models}.
\Secref{Gauss} is devoted to more generally analysing statistical dipole fluctuations at
different Rayleigh numbers which lead us to defining a Gaussian model for reversals.
Based on this model we proceed with quantifying the  properties of reversals and excursions
in \secref{3Stages} and \secref{Statistics}.
A discussion in \secref{Discussion} closes the paper.


%% file: Methods2.tex
\section{Numerical Methods}
\label{Methods}
Since the details of the empoyed numerical dynamo model MagIC3 have been explained
elsewhere \citep{Wicht2002,Christensen2007} we concentrate on outlining
only the essential ingredients here.
A coupled system of equations describing convective motions and
dynamo action is solved in a rotating spherical shell
that represents Earth's outer core.
These equations are the Navier-Stokes equation in the
Boussineq approximation \refp{NS}, the induction equation \refp{Dynamo}, and
a transport equation for codensity $C$ \refp{heat}.
They are supplemented by the simplified continuity equation \refp{divU}
and the condition \refp{divB} that the magnetic field is divergence free:
\begin{equation}
\begin{split}
\label{NS}
\E\,\Pm^{-1}\;\left(\,\dot{\U} + \U\cdot\nabla\U\right)
- \E\;\nabla^2\U +
2\hat{\bf z}\times\U +
\nabla P = \\ \Ra\,\Pm\;r/r_{o} C \hat{r}
+ (\nabla\times\B)\times\B
\end{split}
\end{equation}

\begin{equation}
\label{Dynamo}
\dot{\B} = \nabla\times(\U\times\B)
+ \;\nabla^2 \B
\end{equation}

\begin{equation}
\label{heat}
\dot{C} + \U\cdot\nabla C =\;\Pm\,\Pra^{-1}\nabla^2 C + \epsilon
\end{equation}

\begin{equation}
\label{divB}
\nabla\cdot{\B} = 0
\end{equation}

\begin{equation}
\label{divU}
\nabla\cdot{\U} = 0
\end{equation}

$\B$, $\U$, and $P$ are magnetic induction,
velocity, and pressure perturbation, respectively.
Vectors $\hat{\bf r}$ and $\hat{\bf z}$ denote the unit vectors in radial direction
and along the rotation axis and dots over a variable indicate a partial time derivative.
The codensity $C$ can stand for the super-adiabatic temperature or the
relative contribution of light elements in the outer core.
The spherical shell is bounded by an insulating mantle at $r=r_o$ and an
electrically conducting inner core at $r=r_i$ which rotates
subject to viscous and Lorentz torques \citep{Wicht2002}.

The following scales have been used to make above equations dimensionless:
The shell thickness $d=r_o-r_i$ serves as the length scale and
the magnetic field is scaled by $(\sigma/\rho\Omega)^{1/2}$, with $\rho$ the
outer core density, $\sigma$ the electric conductivity, and $\Omega$ the basic rotation rate.
Time is measured in units of the magnetic diffusion time $t_\lambda=d^{2}/\lambda$,
where $\lambda = (\mu\sigma)^{-1}$, and  $\mu$ is the magnetic permeability.
The dipole decay time is a fraction of the magnetic diffusion time: $\tdip=t_\lambda r_o^2/(d^2\pi^2)$.
The codensity gradient across the outer core, $\Delta C /d$, serves as a codensity unit
in models where $C$ is kept fixed at the boundaries. This is equivalent
to imposing a constant temperature gradient $\Delta T=\Delta C/\alpha$ with
thermal expansivity $\alpha$.
Two setups explore the effects of purely compositional driving from a growing inner core
which we model by setting the codensity flux through the outer boundary to zero. The
sink term $\epsilon$ in \eqnref{heat} is then used to balance the codensity flux from the inner core
and also serves as a codensity scale. Rigid flow boundary conditions are used in all models.

The numerical model then comprises five dimensionless numbers:
The Ekman number $E=\nu/\Omega d^2$,
the (modified) Rayleigh number $Ra=g_o \delta C d / \nu\Omega$,
the Prandtl number $\Pra=\nu/\kappa$,
the magnetic Prandtl number $\Pm=\nu/\lambda$,
and the aspect ratio $a=r_i/r_o$.

%% file: DynamoModels2.tex
\begin{landscape}
\begin{table}
{\scriptsize
\begin{tabular}{c|*{6}{c}|*{13}{c}}
Name & Revs. &\E & \Pm & BC & \Ra & \Ra/\Ra$_c$ & $\Rm$ & $\ElsCMB$ & $\Ro_\ell$
&$\mbox{D}_{geo}$&$\trans$&$\all$&$N_{R\theta}$&$N_{E\theta}$&$N_{e\theta}$ \\
\hline\hline
E3R5 &-& $10^{-3}$   & $10$ & temp. & $250$ & $4.5$ &$202$&  $2.6$&$0.04$&$0.60$& $0.00$&   $137$&0   &0  &0\\
E3R7 &-&             &      &       & $400$ & $7.2$ &$350$&  $2.0$&$0.09$&$0.44$& $0.00$&    $58$&0   &0  &0\\
E3R8 &E&             &      &       & $450$ & $8.1$ &$393$&  $1.8$&$0.11$&$0.38$& $0.01$&    $87$&4   &1  &10\\
E3R9 &E&             &      &       & $500$ & $8.9$ &$436$&  $1.5$&$0.12$&$0.31$& $0.04$&$4\,208$&542 &613&1963\\
E3R13&M&             &      &       & $750$ & $13.4$&$592$&  $1.7$&$0.17$&$0.18$& $0.34$&    $55$&--  &-- &--\\
\hline
E3R9C &-& $10^{-3}$  & $10$ & chem. & $500$ & $9.4$ &$229$&  $2.1$&$0.05$&$0.57$& $0.00$&    $98$&0   &0  &0\\
E3R19C&-&            &      &       & $1000$& $18.8$&$374$&  $1.2$& $0.09$&$0.42$& $0.00$&    $77$&0   &1  &3\\
E3R23C&E&            &      &       & $1250$& $23.5$&$442$&  $0.65$& $0.11$&$0.30$& $0.05$&   $431$&105 &104&234\\
E3R28C&M&            &      &       & $1500$& $28.2$&$491$&  $0.56$& $0.12$&$0.23$& $0.18$&   $323$&--  &-- &--\\
E3R38C&M&            &      &       & $2000$& $37.6$&$564$&  $0.71$& $0.14$&$0.19$& $0.42$&    $45$&--  &-- &--\\
\hline
E4R53C &-& $3\tp{-4}$& $3$  & chem. & $3000$& $80$  &$264$& $0.34$& $0.09$&$0.66$& $0.00$&     $11$&0   &0   &0\\
E4R78C &-&           &      &       & $4500$& $120$ &$340$& $0.25$& $0.11$&$0.59$& $0.02$&    $45$&0   &2   &1 \\
E4R106C&E&           &      &       & $6000$& $160$ &$408$& $0.14$& $0.13$&$0.40$& $0.14$&    $87$&20  &30  &45\\
E4R159C&M&           &      &       & $9000$& $240$ &$497$& $0.15$& $0.16$&$0.27$& $0.10$&    $20$&--  & -- &--\\
\hline
E2R1.8&-& $2\tp{-2}$ & $10$ & temp. & $220$ & $1.8$&$65$ & $0.37$& $0.16$&$0.50$& $0.00$&$1\,342$& 0  &0  &0\\
E2R2.2&E&            &      &       & $260$ & $2.2$&$81$ & $0.20$& $0.20$&$0.38$& $0.05$&$1\,220$&160 &32 &64\\
E2R2.5&E&            &      &       & $300$ & $2.5$&$94$ & $0.33$& $0.23$&$0.40$& $0.06$&$9\,583$&2956&948&2279\\
E2R4.2&M&            &      &       & $500$ & $4.2$&$148$& $0.42$& $0.37$&$0.28$& $0.21$&$346$   &--  &-- &--\\
\hline
Earth &E& $10^{-15}$   &$10^{-6}$& flux    &   --   &  --  &$2000$&$O(1)$&$0.09$&$0.79$& $<0.6$ &$>12\,000$&--&--\\
\end{tabular}
}
\caption{\label{TabPar}
List of parameters and time averaged properties for the explored dynamo models and Earth.
Column 4 details the driving mechanism: an imposed constant temperature jump across the shell (temp.),
or chemical convection (chem.).
The Prandtl number is unity in all cases.
Magnetic Reynolds number $\Rm$, CMB Elsasser number $\ElsCMB$, local Rossby number $\Ro_\ell$, and
relative CMB dipole strength $\Dgeo$ are time averaged values.
Earth values refer to molecular diffusivities, the GUFM field model, and the
local Rossby number estimate by \citet{Christensen2006}.
The relative transitional dipole time $\trans$ in column 12 is
the fraction of time the magnetic north pole spends further away from the
closest geographic pole than $45^\circ$. Column 13 gives the total simulation time in units
of dipole diffusion time and the last three columns list
the number of reversal, grand excursions, and excursions based on the tilt criterion.
See main text for more explanations.}
\end{table}
\end{landscape}

\section{Model overview}
\label{Models}
To study the statistical behaviour of dipole variations and reversals we compiled extremely
long simulations of dynamo models at three different Ekman numbers ($2\tp{-2},10^{-3},3\tp{-4}$) and
different Rayleigh numbers. All models assume a Prandtl number $\Pr=1$
and an Earth-like aspect ratio $a=0.35$.
The magnetic Prandtl number is $\Pm=3$ for the lowest Ekman number models and $10$ otherwise.
Some of these models have previously been discussed in the literature \citep{Kutzner2002,Wicht2005,Wicht2010}.
The naming of the models follows the convention $ExR?$ where $x$ is the unsigned exponent in the Ekman number
and the question mark stands for the supercriticality $\Ra/\Ra_c$, $\Ra_c$ being the critical
Rayleigh number for onset of convection. A $C$ at the end of the name refers to compositional
convection modelled with a vanishing codensity flux at the outer boundary.
The four different model setups E3R?, E3R?C, E4R?C, and E2R? have been explored, gradually increasing
the Rayleigh number until the multipolar regime is reached in each case.
The $C$ at the end of the name refers to purely compositional convection.

\Tabref{TabPar} list the model parameters along with important time averaged non-dimensional characteristics.
In the scaling used here, the non-dimensional magnetic field amplitude is
identical to the Elsasser number, a measure for the ratio of Lorentz to Coriolis forces
in the Navier-Stokes equation:
\bel{Els}
\Els = \frac{\mbox{B}^2 \sigma}{\rho\Omega}\;\;.
\ee
The time averaged outer boundary rms field provides an estimate for
Elsasser number $\ElsCMB$ at the core mantle boundary in \tabref{TabPar}.
The magnetic Reynolds number
\bel{Rm}
\Rm = \frac{\mbox{U} d}{\lambda}
\ee
quantifies the ratio of magnetic field induction to dissipation.
Time averaged rms flow amplitudes U serve to calculate the respective values
listed in \tabref{TabPar} and enter the local Rossby number already defined in
\secref{Intro}.
Another listed dimensionless property is the dipolarity \Dgeo, the ratio of the rms dipole
field at the CMB to the rms field based on all spherical harmonic
contributions up to degree and order $11$.

\begin{figure}
\begin{center}
\includegraphics[width=11cm]{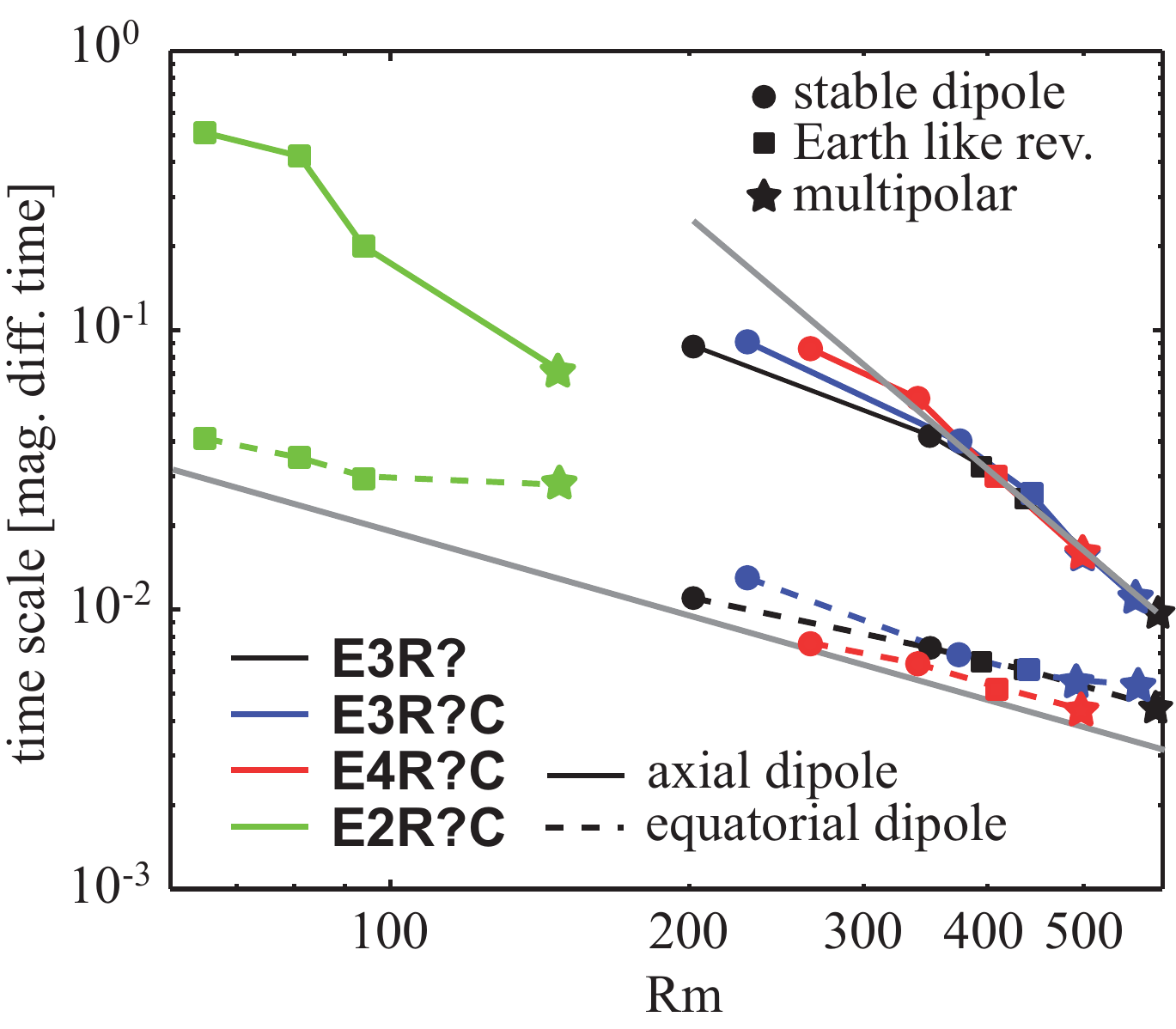}
\end{center}
\caption{Dependence of typical time scales for the axial (solid lines) and
equatorial dipole contribution (dashed) in all the models explored here. The lower grey line
illustrates the $\Rm^{-1}$ scaling suggested by \citet{Christensen2004}.
The upper grey line scales like $\Rm^{-3}$, a slope purely indicated by the data.
Symbol types code the reversal behaviour.}
\label{TauRm}
\end{figure}

\Figref{TauRm} provides an overview of the different models in terms of the typical dipole time scales and
the reversal behaviour. Following \citep{Hulot1994} and \citep{Christensen2004}
we define the time scale of a specific rms magnetic field contribution
$B_{\ell m}$ by
\bel{TS}
\tsv_{\ell m}=\left( \frac{\ovl{{B}^2_{\ell m}}}
{\ovl{\left( B_{\ell m}(t+\delta t)-B_{\ell m}(t)\right)^2 / \delta t^2 }}\right)^{1/2}\;\;
\ee
where the overline stands for the time average.
The time scales $\tsv_{10}$ and $\tsv_{11}$ of the axial and equatorial dipole both decrease with
increasing Rayleigh number. The former, however, is typically an order of magnitude larger than the latter.
\citet{Christensen2004} and \citet{Lhuillier2011} report that the time scale for higher multipoles $\ell>1$
is inversely proportional to the spherical harmonic degree and also the magnetic Reynolds number,
\bel{Tau}
\tsv_{\ell}\sim (\ell\Rm)^{-1}\;\;,
\ee
while the dipole contribution has a significantly slower time scale.
\Figref{TauRm} demonstrates that this is mainly a consequence of the slow axial dipole
variations (solid lines) while  the equatorial dipole contribution (dashed lines) roughly
agrees with the time scale $\tsv_1$ suggested by \eqnref{Tau} (lower grey line).
When increasing the Rayleigh number and thus $\Rm$, the axial dipole progressively
looses its stability. Being very stable at low \Ra\ (circles in \figref{TauRm}),
it starts to show Earth-like rare reversals when \Ra\ is sufficiently high (squares) and ultimately enters a
multipolar regime where the axial dipolar looses its dominance and reverses
more or less continuously (stars).
Somewhat arbitrarily, we regard reversals as Earth-like when
the relative transitional time $\trans$ the magnetic pole spends between plus and minus $45^\circ$
latitude is not larger than $0.15$ and the mean dipolarity $\Dgeo$ exceeds $0.3$.
Similar but slightly stricter definitions have been used by \citet{Wicht2005} and \citet{Wicht2010b}.

Once reversals have started, the time scale of the axial dipole $\tau_{10}$ decreases
rapidly with a slope close to $\Rm^{-3}$, as is demonstrated in \figref{TauRm}, and thus progressively
approaches the faster time scale of the equatorial dipole and higher multipole contributions.
This further confirms that the axial dipole looses its special role in strongly driven systems.
Models E2R?C have not only a particularly large Ekman number but also a much smaller magnetic
Reynolds number than the other cases.
\Figref{TauRm} indicates that this results in a very different behaviour. The
dynamo can already reverse very close to the onset of dynamo action but with a very different type of reversals
than at smaller Ekman numbers, as we will further discuss below.
Another difference is that the local Rossby number already significantly exceeds the critical value
$\Ro_\ell\approx 0.1$ around which reversals can be expected.

\begin{figure}
\centering
\includegraphics[draft=false,width=10cm]{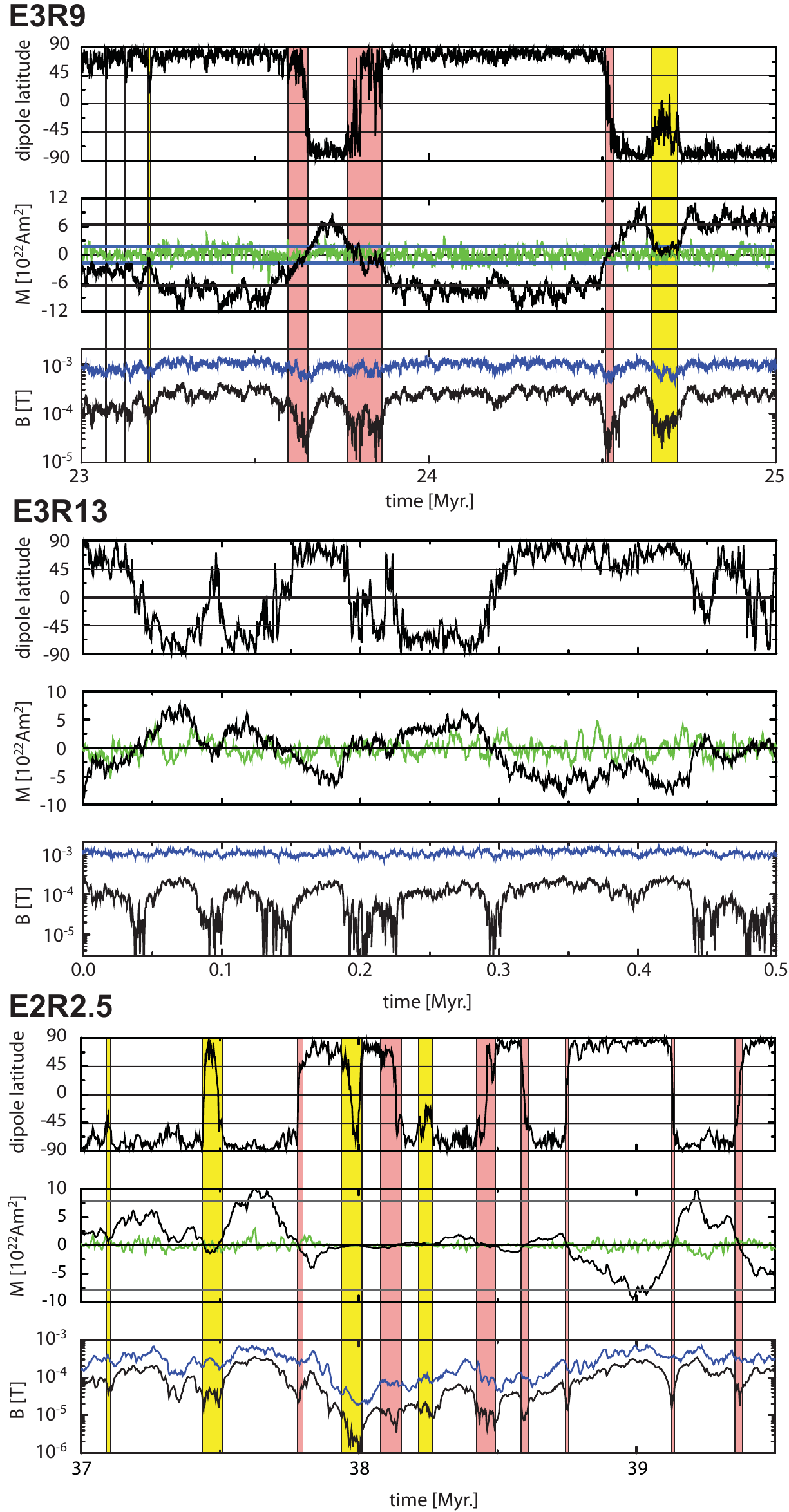}
\caption{Examples for the evolution of different field characteristics in three dipole models.
Upper sub-panels show the dipole tilt, middle sub-panels show axial dipole moment
(black) and real equatorial dipole moment (green), and the lower sub-panels
show the rms field strength at the CMB due to the axial dipole moment (black) and due to
higher harmonic contributions $\ell>1$. Thick grey horizontal lines in the middle sub-panels
show estimates for the mean axial dipole moment of the high dipole moment state while the horizontal
blue lines show (eye-ball) estimates for the mean ADM in a lower dipole moment state. For the Earth-like
reversing model we have indicated transitional periods with $\theta>45$ with red and yellow
background colours for reversals and excursion, respectively.}
\label{RStimeADM}
\end{figure}

\Figref{RStimeADM} illustrates the temporal evolution for three selected models.
The top sub-panel for each model depicts the dipole latitude
$\theta=\arctan{(B_{11}/B_{10})}$ given in degree.
Middle panels show axial dipole moment (AMD) and the real equatorial dipole moment (EDM).
Since there is no preferred longitude in our models both the real and imaginary
(or $\sin$ and $\cos$) contributions of the equatorial dipole are interchangeable and
show exactly the same (statistical) behaviour. This would be different if the longitudinal symmetry
would be broken by, for example, imposing a non-axisymmetric heat flux at the outer boundary condition.
Vertical coloured background stripes in \figref{RStimeADM} mark periods during which the
dipole tilt angle $\Theta=90^\circ-|\theta|$ exceeds $45^\circ$, reversals in red and excursions in yellow.
Following \citet{Wicht2009} we neglect any events shorter than $2\,$kyr and melt events closer
together than $40\,$kyr to form one longer lasting event. The melting takes into
account that it requires some time to reestablish the stable polarity epoches that separate
individual events.
When the stable polarity epochs before and after an event have the same polarity, the
event is classified as an excursion rather than a reversal.
We also distinguish a particular subset of excursions where the dipole ventures into
the opposite polarity. These deeper 'grand excursions' are more likely to be detected
globally in paleomagnetic records \citep{Wicht2009}.
The last three columns in \tabref{TabPar} list the number of reversals $N_{R\theta}$,
grand excursions $N_{E\theta}$, and excursions $N_{e\theta}$ detected in each of the dynamo runs.
We will for simplicity refer to this set of conditions as the
tilt criterion in the following and mark respective quantities with the subscript $\theta$.

The axial dipole moment changes on many different time scales and shows a very complex behaviour.
In the Earth-like reversing model E3R9 shown in \figref{RStimeADM} we can distinguish a high dipole
moment state HDM and a low dipole moment state LDM. As will be explained below, the mean ADM in the high dipole
moment state $\overline{M}_{H10}=7.1\times10^{22}$Am$^2$ can be estimated
based on the statistics (thick grey horizontal line) while the LDM mean $\overline{M}_{L10}=1.7\times10^{22}$Am$^2$
(horizontal blue line) is merely a tentative suggestion.
During reversals and the longer lasting deeper excursions, the ADM tends
to remain in the LDM state for some time before switching back to the HDM configuration.
These events are thus not simple transitional between two states of opposite polarity
but involve a new weak dipole moment state.
Though a closer inspection of the ADM variation reveals more complex details, the two state
description seems appropriate for characterizing the reversal behaviour, as will be demonstrated below.
The equatorial dipole moment contributions show a much simpler variation
around a zero mean with on average shorter time scales (see middle panels in \figref{RStimeADM}, green lines).

At lower Rayleigh numbers before the onset of any reversals (not shown), ADM and EDM variations show similar behaviour
but the ADM simply oscillates around a larger mean of one sign and the LDM state is missing.
As \Ra\ is increased beyond the transition to the multipolar regime, reversals and excursions become
much more frequent and a separation between individual events or HDM and LDM becomes difficult,
as is illustrated with model E3R13 in \figref{RStimeADM}.
The ADM is significantly weaker so that the dipole
has lost its dominance and the magnetic field appears to be multipolar or complex. ADM variations
still clearly show longer time scales than the EDM contributions though the difference is
not as pronounced as in E3R9 (see \figref{TauRm}).

The lower sub-panel for model E3R9 in \figref{RStimeADM} reveals a high degree of correlation
between the total rms dipole field and the dipole tilt angle.
The correlation between the multipole field contribution and larger tilt angles is still
detectable but much weaker. \Tabref{TabC} list various correlation coefficients for all dynamo models.
For mild Rayleigh numbers where dipole tilts remain generally small, the tilt variations
are mainly caused by equatorial dipole variations as attested by the
high correlation coefficients $\corr(\mbox{B}_{11},\Theta)\ge0.8$.
When the Rayleigh number is increased, axial dipole fluctuations start to play a more
important role and the respective correlation coefficient $\corr(\mbox{B}_{10},\Theta)$
increases up to $0.7$ for reversing models while $\corr(\mbox{B}_{11},\Theta)$ decreases.
Large tilt angles are generally caused by a decreasing axial dipole contribution.
The smaller but still significant correlation between tilt angle and higher multipoles
is a consequence of the high correlation between axial dipole and multipole contributions
$\corr(\mbox{B}_{10},\mbox{B}_{\ell>0})$ which typically exceeds $0.8$.
The high value reflects the coupling of different scales via the dynamo process
where the dipole contributions is produced from higher degree contributions and vice versa.
The correlation between the equatorial dipole and higher harmonics is significantly
lower likely because the equatorial dipole belongs to the less preferred equatorially
symmetric dynamo family.
In the multipolar regime, $\corr(\mbox{B}_{10},\mbox{B}_{\ell>0})$ decreases while
$\corr(\mbox{B}_{11},\mbox{B}_{\ell>0})$ remains largely unchanged, at least for the
smaller Ekman number dynamos \citep{Wicht2010b}. This is another indication that the axial dipole
looses its special role at larger Rayleigh numbers.

\begin{table}
\begin{tabular}{c|*{6}{c}}
Name & Revs. &$\corr(\mbox{B}_{10},\Theta)$&$\corr(\mbox{B}_{11},\Theta)$
             &$\corr(\Dgeo,\Theta)$&$\corr(\mbox{B}_{10},\mbox{B}_{\ell>0})$
                                   &$\corr(\mbox{B}_{11},\mbox{B}_{\ell>0})$\\
\hline\hline
E3R5 &-& $ 0.05$&$ 0.96$&$ 0.08$&$0.72$&$0.26$\\
E3R7 &-& $-0.20$&$ 0.85$&$-0.15$&$0.83$&$0.30$\\
E3R8 &E& $-0.51$&$ 0.48$&$-0.54$&$0.85$&$0.32$\\
E3R9 &E& $-0.62$&$ 0.39$&$-0.61$&$0.83$&$0.34$\\
E3R13&M& $-0.71$&$ 0.43$&$-0.53$&$0.53$&$0.33$\\
\hline
E3R9C &-&$-0.11$&$ 0.98$&$ 0.03$&$0.55$&$0.15$\\
E3R19C&-&$-0.13$&$ 0.75$&$-0.14$&$0.87$&$0.47$\\
E3R23C&E&$-0.61$&$ 0.28$&$-0.62$&$0.86$&$0.42$\\
E3R28C&M&$-0.68$&$ 0.37$&$-0.58$&$0.71$&$0.37$\\
E3R38C&M&$-0.72$&$ 0.44$&$-0.46$&$0.48$&$0.40$\\
\hline
E4R53C &-&$-0.09$&$0.97$&$-0.07$&$0.77$&$0.15$\\
E4R78C &-&$-0.35$&$0.64$&$-0.47$&$0.88$&$0.31$\\
E4R106C&E&$-0.63$&$0.20$&$-0.70$&$0.87$&$0.39$\\
E4R159C&M&$-0.68$&$0.33$&$-0.66$&$0.70$&$0.34$\\
\hline
E2R1.8&-&$-0.02$&$ 0.80$&$ 0.09$&$0.83$&$0.10$\\
E2R2.2&E&$-0.45$&$-0.17$&$-0.63$&$0.97$&$0.73$\\
E2R2.5&E&$-0.59$&$ 0.13$&$-0.65$&$0.88$&$0.46$\\
E2R4.2&M&$-0.61$&$ 0.24$&$-0.42$&$0.80$&$0.57$\\
\end{tabular}
\caption{\label{TabC}
Correlation coefficients between various rms field contributions, the tilt angle $\theta$,
and the dipolarity \Dgeo.}
\end{table}



The other two lower Ekman number setups E3R?C and E4R?C show very similar behavior but setup
E2R? is clearly different.
\Figref{RStimeADM} demonstrates that the dynamo stops operating intermittently in model E2R2.5
so that all field components decay and only slowly recover some time later. The
axial dipole may still dominate during these epochs and undergo stochastic reversals and excursions.
However, the total field can assume a rather low level and the dipole remains relatively unstable
even in the inter-event periods.
We will refer to these special events as reversals and excursions of type 2 to distinguish
them from the more typical events of type 1 which predominantly concern variations in the axial dipole component.
Model E2R2.5 shows reversals of both types while the smaller \Ra\ in models E2R1.8 and E2R2.2
allows only for reversals of type 2. Another difference between the large and the lower Ekman
number models is that the Earth-like reversing case E2R2.5 shows more reversals than excursions (see \tabref{TabPar}).

\section{A Gaussian reversal model}
\label{Gauss}

\begin{figure}
\centering
\includegraphics[draft=false,width=12cm]{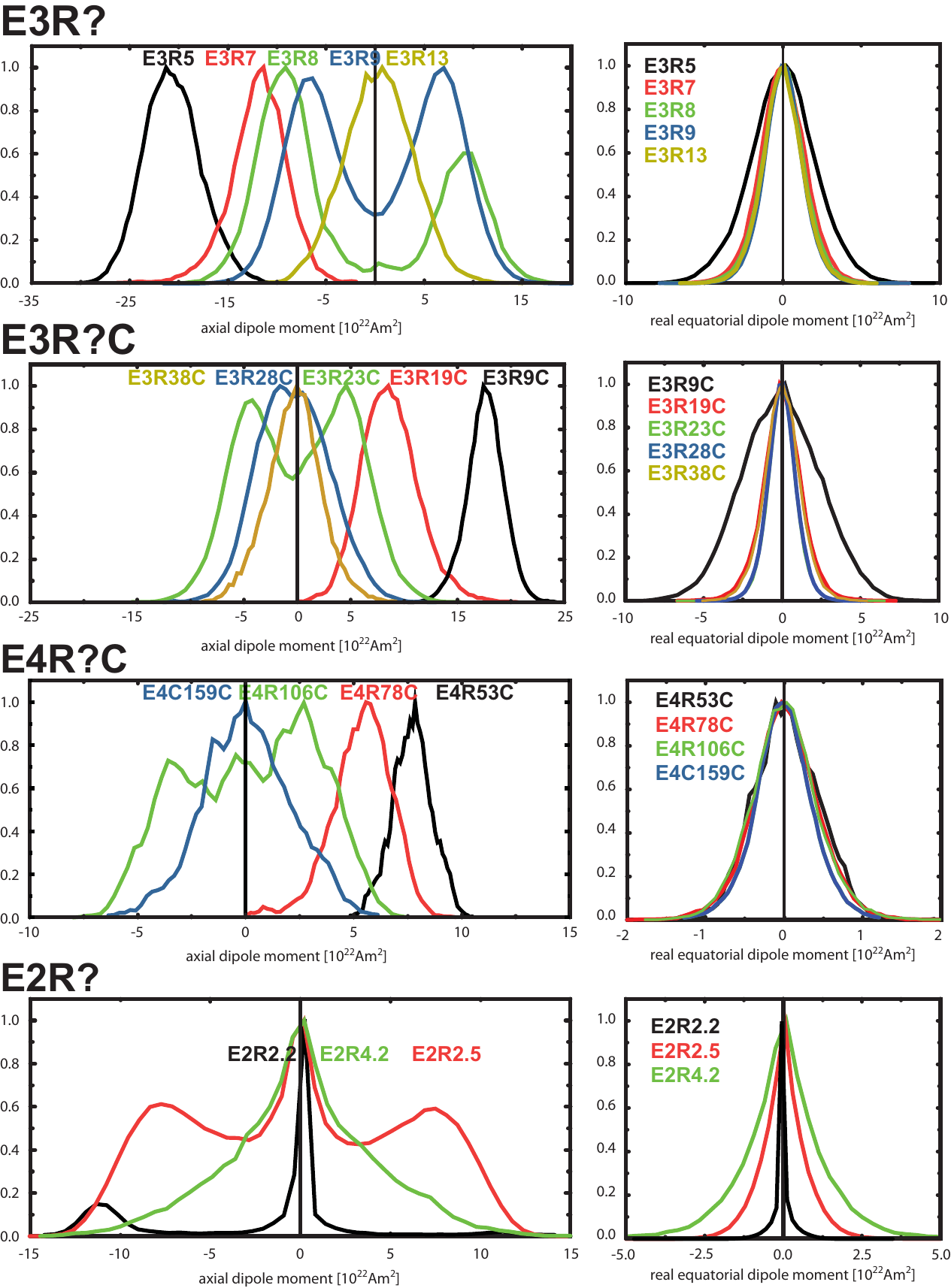}
\caption{Rayleigh number dependence of axial dipole moment histograms (left)
and real equatorial dipole moment histograms (right) for all explored models.
All histograms have been normalized with the respective maximum.}
\label{DM_all_bins}
\end{figure}

\Figref{DM_all_bins} shows how the histograms of axial and real
equatorial dipole moment change with Rayleigh number in the different model setups.
We start with discussing the rather similar properties in models E3R?, E3R?C, and E4C.
The ADM assumes a rather simple Gaussian shape for smaller Rayleigh numbers.
On increasing \Ra\ the histogram retains its overall shape while
the (absolute) mean decreases. Reversals start once low dipole moments become more likely and
the histogram then assumes a bimodal shape with ADM values of both signs.
However, the low ADM states are more populated than a superposition of two symmetric Gaussians would imply.
The relative likelihood of low values further increases with \Ra\ until the distribution
looks like a Gaussian with zero mean in the multipolar regime.

While \figref{RStimeADM} seemed to indicate that the low ADM state may also consist of
two symmetric distributions this is not supported by the AMD histograms.
However, the superposition of the two Gaussians may simply be indistinguishable from
a distribution with zero mean and larger standard deviation.
Similarly, the two high ADM states may still contribute
to the flanks of the distribution in the multipolar regime.
The EDM histograms in the lower Ekman number cases remain surprisingly similar for all Rayleigh numbers and
always assume a Gaussian shape with zero mean (see \figref{DM_all_bins}, right column).

Several authors have suggested that the geomagnetic field actually obeys a Gaussian model with
a non-zero mean for the axial dipole and zero means for all other field coefficients \citep{Constable1988,Hulot1994}.
Our simulations suggest that this model oversimplifies the axial dipole distribution.
The required correction, however, is likely only small and concerns only data
covering a fair portion of transitional epochs.

\begin{figure}
\centering
\includegraphics[draft=false,width=12cm]{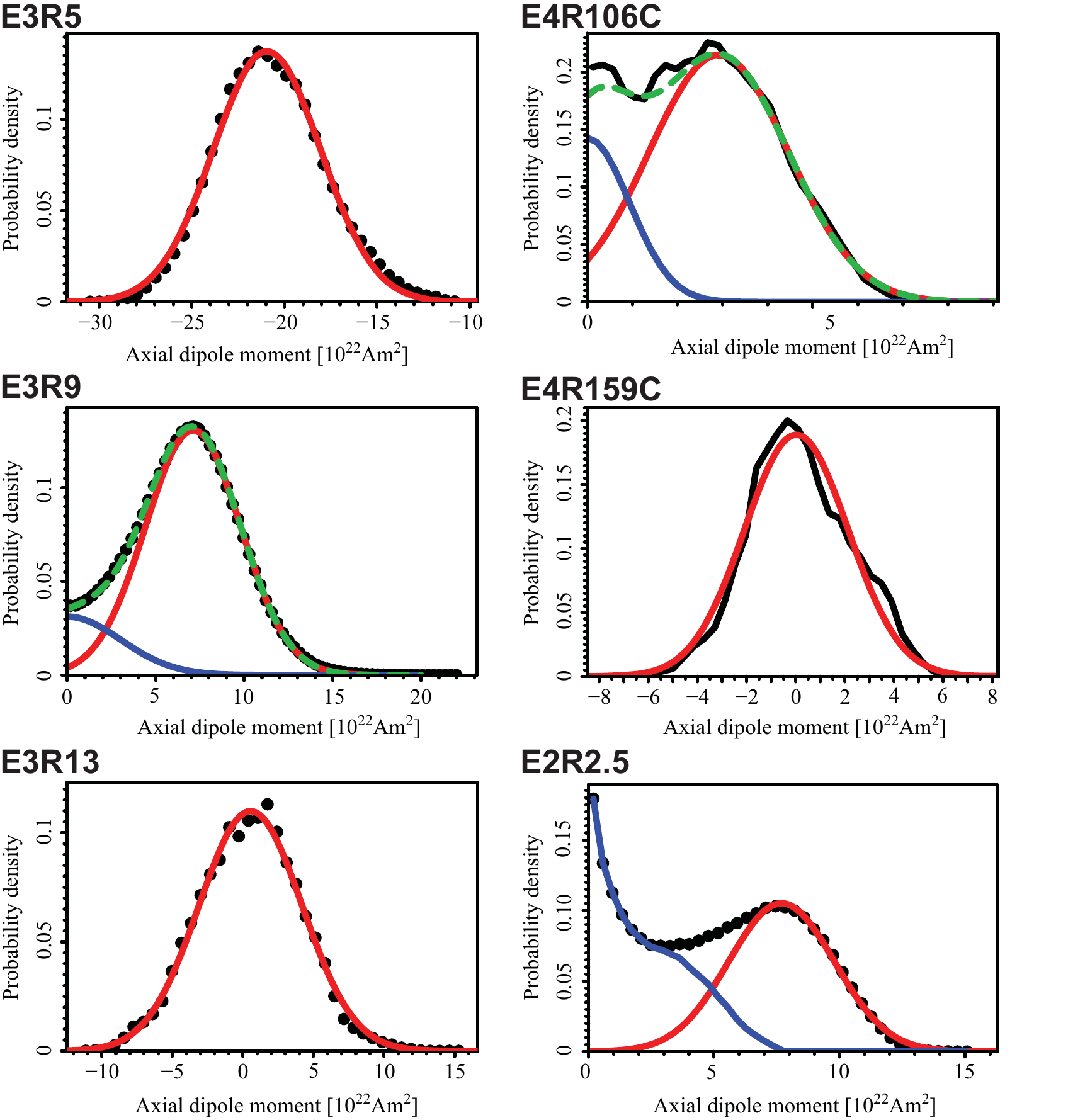}
\caption{Fit of Gauss distributions to some of the ADM histograms. For the
reversing models E3R9 and E4R106C the red and blue lines show the Gauss distributions
for the high and low dipole moment states, respectively. For case E2R2.5 the blue curve
is the difference between the numerical distribution and the red Gauss distribution which was fitted
to the right flank of the numerical distribution.}
\label{GaussFitADMall}
\end{figure}

We have tried to fit Gaussian distributions to the ADM and the EDM components in all models,
using a least-squares procedure for the binned data. The respective results are illustrated for
a few selected cases in \figref{GaussFitADMall}.
To disentangle the HDM and LDM contributions in the Earth-like reversing cases we
analyse the unsigned dipole moment,
assuming that the distributions should be symmetric around zero.
The HDM axial dipole Gaussian is then fitted to the right flank of the unsigned ADM
distribution beyond the distribution maximum.
The LDM Gaussian is fitted in a second step after subtracting the HDM model.
The Gaussians convincingly reproduce the ADM and EDM distributions for
models E3R? and E3R?C. The histograms are less smooth for the shorter runs of E4R?C
but the model distributions nevertheless seem to provide a fair representation, as is demonstrated
in \figref{GaussFitADMall}.

The histograms for E2R? clearly deviate from the Gaussian models and reflect the intermittent nature
of the dynamo process in the form of pronounced peaks around zero.
Here we only fitted the HDM part of the axial dipole moment.
The peaked equatorial dipole moment, LDM axial dipole moment,
and multipolar axial dipole moment distributions are far from a normal distribution.

\begin{figure}
\centering
\includegraphics[draft=false,width=12cm]{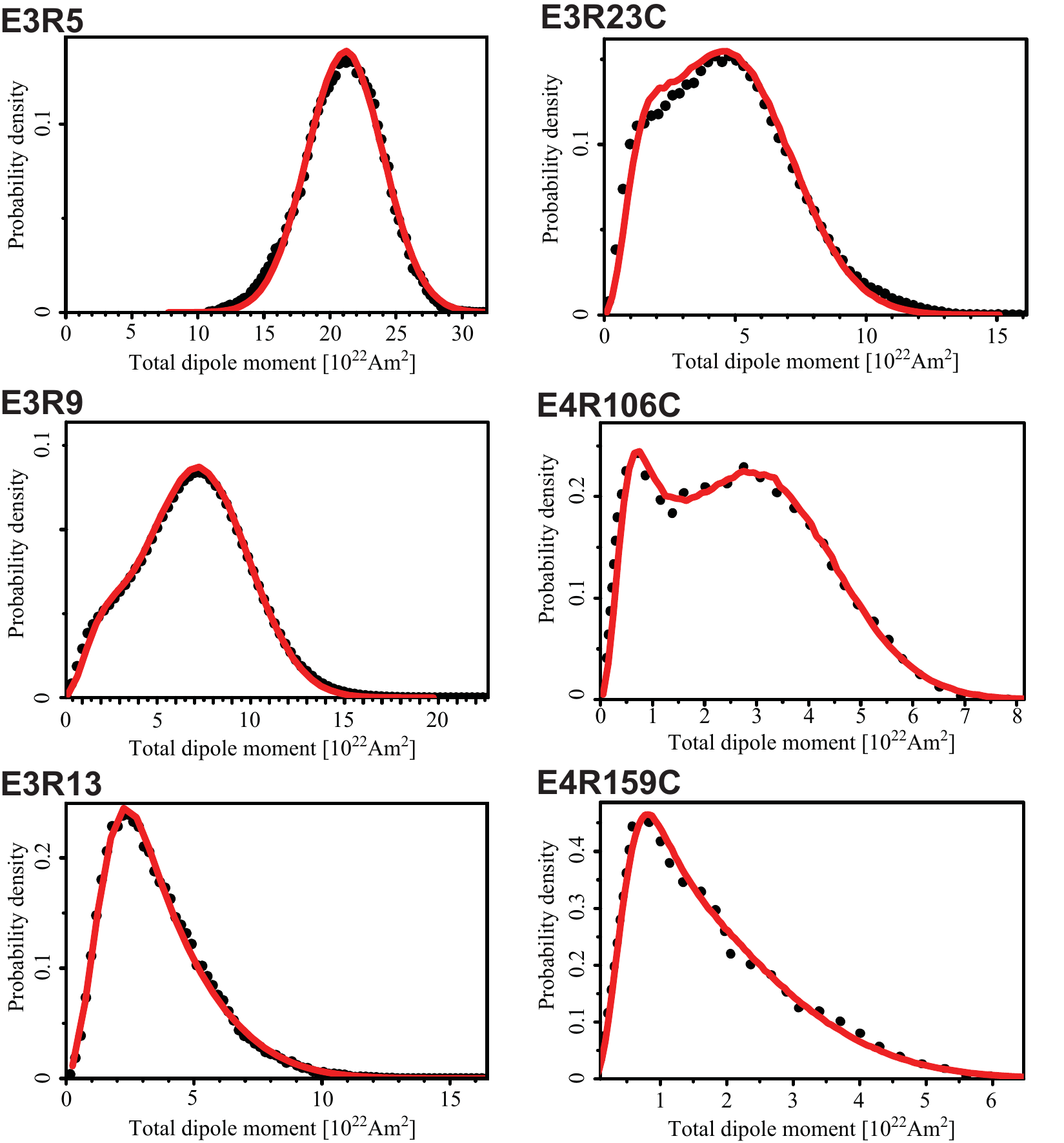}
\caption{Histograms of the total dipole moment in different dynamo models. Black dots show the numerical
simulation data while red lines are the predictions based on drawing random samples from
the axial and equatorial dipole moment distributions.}
\label{DMpred}
\end{figure}

\begin{figure}
\centering
\includegraphics[draft=false,width=12cm]{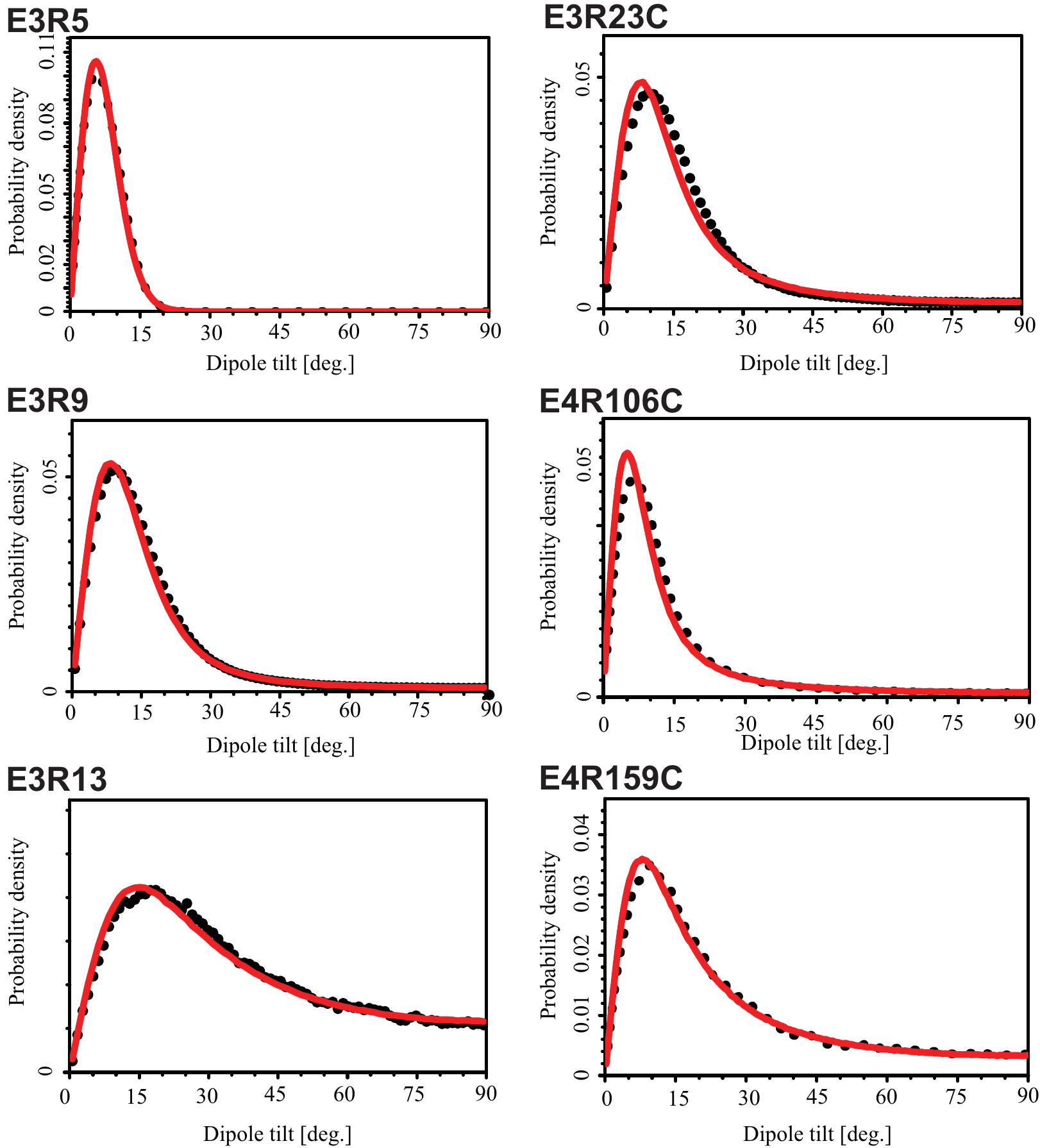}
\caption{Histograms of the dipole tilt angle in different dynamo models. Black dots show the numerical
simulation data while red lines are the predictions based on the Gaussian models.}
\label{TiltPred}
\end{figure}

The correlation coefficients between all three dipole contributions
never exceed a few percent so that they are likely statistically independent.
This suggests that the total dipole moment $M$ and the dipole tilt $\theta$ can be predicted
by combining randomly and independently drawn samples from the individual
moment distributions. The procedure indeed produces convincing representations
of the respective numerical distributions as is demonstrated for a few examples
in \figref{DMpred} and \figref{TiltPred}.
The procedure works best for the longer runs E3R? and somewhat less
convincingly for E3R?C and E4R?C. We did not attempt predictions for the larger
Ekman number models E2E? because of their complex non-Gaussian histograms.

\Tabref{TabGauss} list mean and standard deviations for the different Gaussian distributions describing
the dipole moment histograms.
For the smaller Ekman number cases the total dipole moment in the HDM state
can be approximated by the respective axial dipole moment distribution.
For estimating the total dipole moment distribution in the LDM state,
we used randomly drawn samples from the respective axial and
equatorial dipole distributions to first estimate the dipole energy distribution which
closely resembles the expected $\chi^2$ distribution for three independent variables \citep{Hulot1994}.
From this we calculated the mean $\ovl{M}_L$ and its standard deviation listed in \tabref{TabGauss}.
Earth values given in \tabref{TabGauss} are only very rough estimates based on data covering
the last $2\,$Myr by \citet{Valet2005} and \citet{Channell2009}. Note, however, that the recent dipole moment estimates
by \citet{Ziegler2011} yield significantly lower dipole moment amplitudes.

Knowing mean and standard deviation $\sigma$ of the individual Gaussian distributions is
apparently sufficient to predict the dipole moment and tilt probability for the lower
Ekman number simulations with some confidence. The respective values are listed in \tabref{TabGauss}
for the reversing dynamo models that we will further explore below.
Since the two equatorial dipole moments behave
identically we have to specify only two distributions for non-reversing
and three for reversing dynamos. When taking into account that equatorial moment and LDM distributions
have a zero mean, the total dipole moment and tilt predictions
rely on only three parameters for non reversing dipolar dynamos
($\ovl{M}_{10}$,$\sigma(M_{10})$,$\sigma(M_{11})$),
on four parameters for reversing dynamo
($\ovl{M}_{H10}$,$\sigma(M_{H10})$,$\sigma(M_{L10})$,$\sigma(M_{11})$),
and on two parameters for multipolar dynamos ($\sigma(M_{10})$,$\sigma(M_{11})$).
Indices refer to high (H) or low (L) dipole moment states and axial (10) or
equatorial (11) contributions.

\begin{table}
{\scriptsize
\centering
\begin{tabular}{c|*{11}{c}}
Model &$\ovl{M}$&$\sigma(M)$&$\ovl{M}_{H10}$&$\sigma(M_{H10})$&$\ovl{M}_{L10}$&$\sigma(M_{L10})$&
$\ovl{M}_{11R}$&$\sigma(M_{11R})$&$\ovl{M}_{L}$&$\sigma(M_{L})$\\
\hline\hline
E3R9   &$6.6$&$2.9$ &$7.1$ &$2.7$&$0.0$&$3.0$ &$0.0$&$1.3$ &$3.1$&$1.6$\\
E3R23C &$4.6$&$2.4$ &$4.8$ &$2.4$&$0.0$&$2.2$ &$0.0$&$0.84$&$2.2$&$1.2$\\
E4R106C&$2.8$&$1.6$ &$2.9$ &$1.5$&$0.0$&$0.94$&$0.0$&$0.34$&$0.92$&$0.50$\\
E2R2.5 &$5.3$&$3.2$ &$7.7$ &$2.1$&$0.0$&$2.8$ &$0.0$&$0.78$&  -  & - \\
\hline
Earth  &$7$  & -    &$9$   &$3$  & -   & -    & -   &  -   & $2$ & - \\

\end{tabular}
\caption{Mean and standard deviation of the total dipole moment $M$
and the individual  dipole moment contributions in the high dipole moment (H) and low dipole moment (L) states.
Suffix $10$ and $11R$ indicate the axial contribution and the real part of the equatorial contributions, respectively.
High and low state values refer to the fitted Gaussian distributions. The high state total moment $\ovl{M}_H$
is very similar to $\ovl{M}_{H10}$ and is therefore not listed. The low state total moment $\ovl{M}_L$
has been estimated by combining randomly drawn samples from the distributions of ${M}_{L10}$ and ${M}_{11R}$.}
\label{TabGauss}
}
\end{table}

\section{Reversals, a process in three stages}
\label{3Stages}
The two-state scenario outlined above suggests that reversals can be separated into three stages.
The first stage is the decay of the axial dipole moment until the LDM state is reached.
Total dipole moment $M$ or mean field strength also clearly drop since the axial dipole dominates initially.
The period spent in the LDM state defines the second stage during which
ADM variations cause one or several polarity changes until the axial dipole moment grows enough to
re-establish the HDM state in the final third stage.
Since the number of polarity switches during stage two seems to be stochastic it
is simply a matter of chance which axial dipole direction is amplified in stage three.
There should thus be a class of grand excursions which shares the statistical properties
of reversals. To explore this idea we refine our definition of grand excursions, requiring not only
that the dipole ventures into opposite polarity but also assumes amplitudes typical for the LDM state.

To verify and further explore the three-stage model we integrated four
Earth-like reversing models long enough to provide some statistics of
the complex and highly variable processes (see \tabref{TabPar}).
For defining the stages we rely on the total rather than only the axial dipole moment
since only the former is reasonably accessible for palaeomagnetic studies.
The reversals and excursions detected with the tilt criterion serve as a first guess for
determining the start of each event $t_s$ and its end $t_e$ when the DM decreases
below or exceeds a threshold $M_{Hc}$. The event duration
including all three stages is then given by $\tdur=t_e-t_s$.
To determine the decay and growth rate in stage one and three we assume that the decay ends at
$t_{De}$ when the DM decreases below a second threshold $M_{Lc}$ and
the growth stage starts at $t_{Gs}$ once $M_{Lc}$ is once more exceeded. The decay and growth time scales
$\dec$ and $\gro$ are then expressed in fractions
of the dipole decay time by comparing $e$-folding times:
\bel{taud}
 \dec=\frac{t_{De}-t_s}{T_D}\,\frac{M_{Hc}(1-1/e}{M_{Hc}-M_{Lc}}=\frac{\tdec}{\tdip}\;\;,
\ee
\bel{taug}
 \gro=\frac{t_e-t_{Gs}}{T_D}\,\frac{M_{Hc}(1-1/e}{M_{Hc}-M_{Lc}}=\frac{\tgro}{\tdip}\;\;.
\ee
The time spent in the LDM state can be estimated by $\tlev=t_{Gs}-t_{De}$. We will also briefly discuss
the waiting time between events $\twai=t_s^{(n)}-t_e^{(n-1)}$
where the upper index refers to the event number.

To exclude the influence of statistical shorter term fluctuations we have used a running mean of the dipole
moment with a window of $5\,$kyr or $0.1$ dipole decay times. This is about
five times the flow overturn time in the reversing simulations.
Experiments show that windows of a few thousand years provided the most consistent timing of reversals and excursions.
Overlapping consecutive events are simply dismissed which generally reduces the number of reversals $N_{R}$ and
grand excursions $N_{E}$ entering the analysis.

\begin{figure}
\centering
\includegraphics[draft=false,width=14cm]{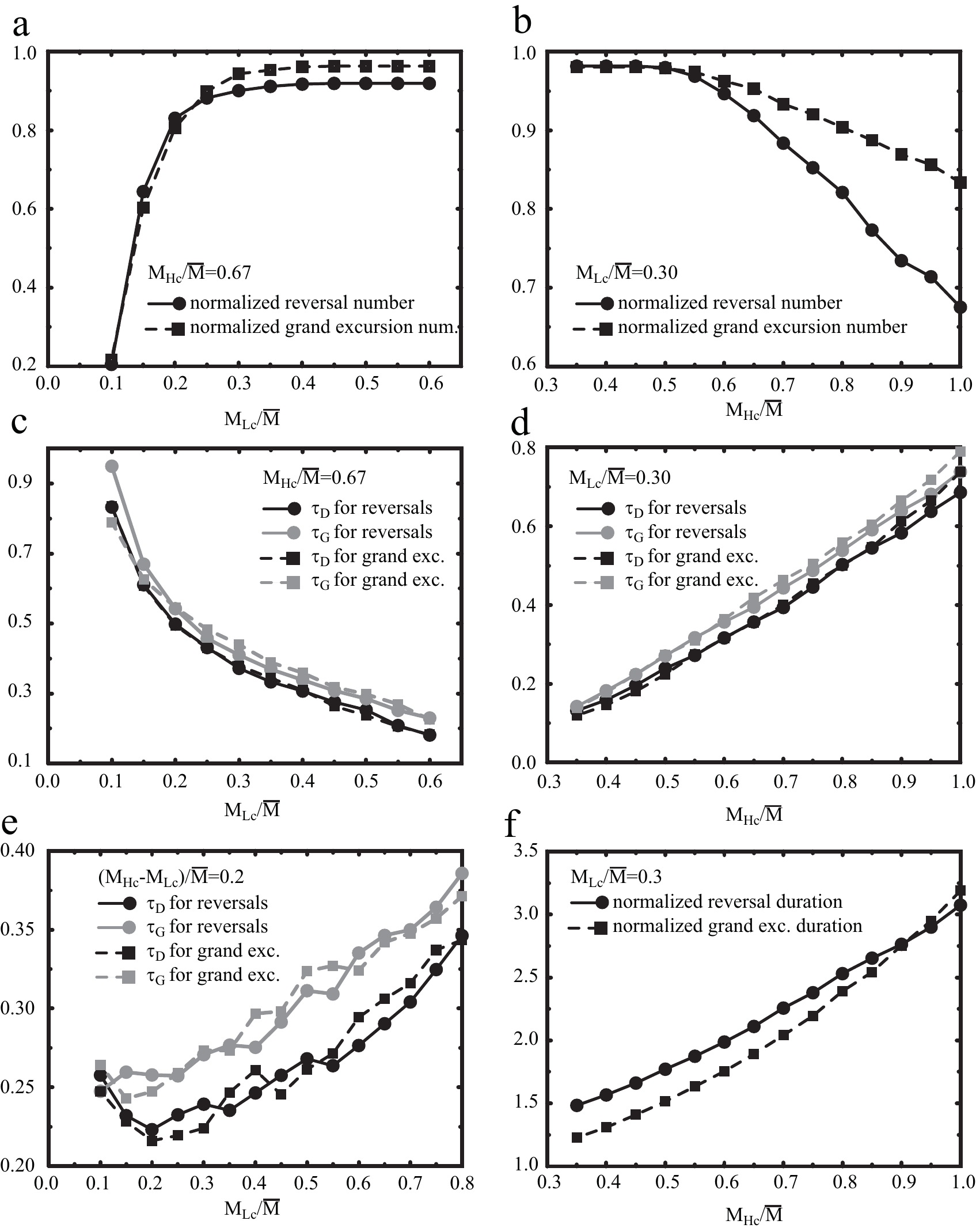}
\caption{Dependence of the number of reversals and grand excursions (a and b) and the
time scales $\tau_D$ and $\tau_G$ (c, d) in model E3R9 on $M_{Hc}$ and $M_{Lc}$ when
the respective other parameter is held fixed. The event number has been normalized with
the event number based in the tilt criterion alone. Panel e illustrates the
dependence of $\tau_D$ and $\tau_G$ on $M_{Lc}$ for a constant difference $M_{Hc}-M_{Lc}$.
Panel f shows the dependence of the event durations on $M_{Hc}$.
}
\label{NrevE3}
\end{figure}

How should the thresholds $M_{Hc}$ and $M_{Lc}$ be chosen? The mean of the HDM state minus its
standard deviation seems a natural first guess for $M_{Hc}$. Since the ADM is much larger than the EDM we can
for simplicity use the ADM mean and standard deviation: $M_{Hc}=\ovl{M}_{H10}-\sigma(M_{H10})$.
Choosing $M_{Lc}=\ovl{M}_L$ seems a reasonable first guess for the lower threshold.
For model E3R9 the above reasoning suggests $M_{Hc}=4.4\tp{22}$Am$^2$
and $M_{Lc}=3.1\times{22}$Am$^2$.
In the following a calligraphic $\MC$ will refer to dipole moments normalized
with the time averaged mean. The E3R9 threshold estimates are then $\MCH=0.67$
and $\MCL=0.47$ respectively.

To understand the role of the thresholds we have explored their impact on the different
reversal and excursion properties.
\Figref{NrevE3} illustrates how the event number and time scales depend
on the choice of $\MCH$ and $\MCL$  for model E3R9.
Panel a reveals that the dipole moment has to decrease to a critical value
of about $30$\% of its mean to facilitate reversals and grand excursions.
For lower values we start losing events where the running mean never decays as much.
If, on the other hand, $\MCH$ exceeds about $0.65$ we disregard
stable polarity intervals during which the dipole moment never exceeds the threshold (see \figref{NrevE3}b).
Considering values outside the interval $(0.30 - 0.65)$ would thus potentially distort
the statistics. Naturally, missing events have a particularly large effect
on the event waiting times (or inter-event times) as will be illustrated below.

Ruling out the statistical short term variations around the
high and low dipole moment states seems essential for estimating the decay
and grow time scales $\dec$ and $\gro$.
This is not an easy task since dipole moment variations are generally complex and
the distributions of HDM and LDM states overlap.
Considering the $5\,$kyr running mean only partially solves this problem and
the event timing remains very sensitive to the threshold values as is demonstrated
in the lower four panels of \figref{NrevE3}.
Asymptotical values of $\dec$ and $\gro$ are only approached when $\MCL$ and $\MCH$
become very similar. The timescales then characterize a rather brief episode
more typical for the faster statistical variations than the slower longer term
transition from the high to the low dipole moment state.

\begin{figure}
\centering
\includegraphics[draft=false,width=12cm]{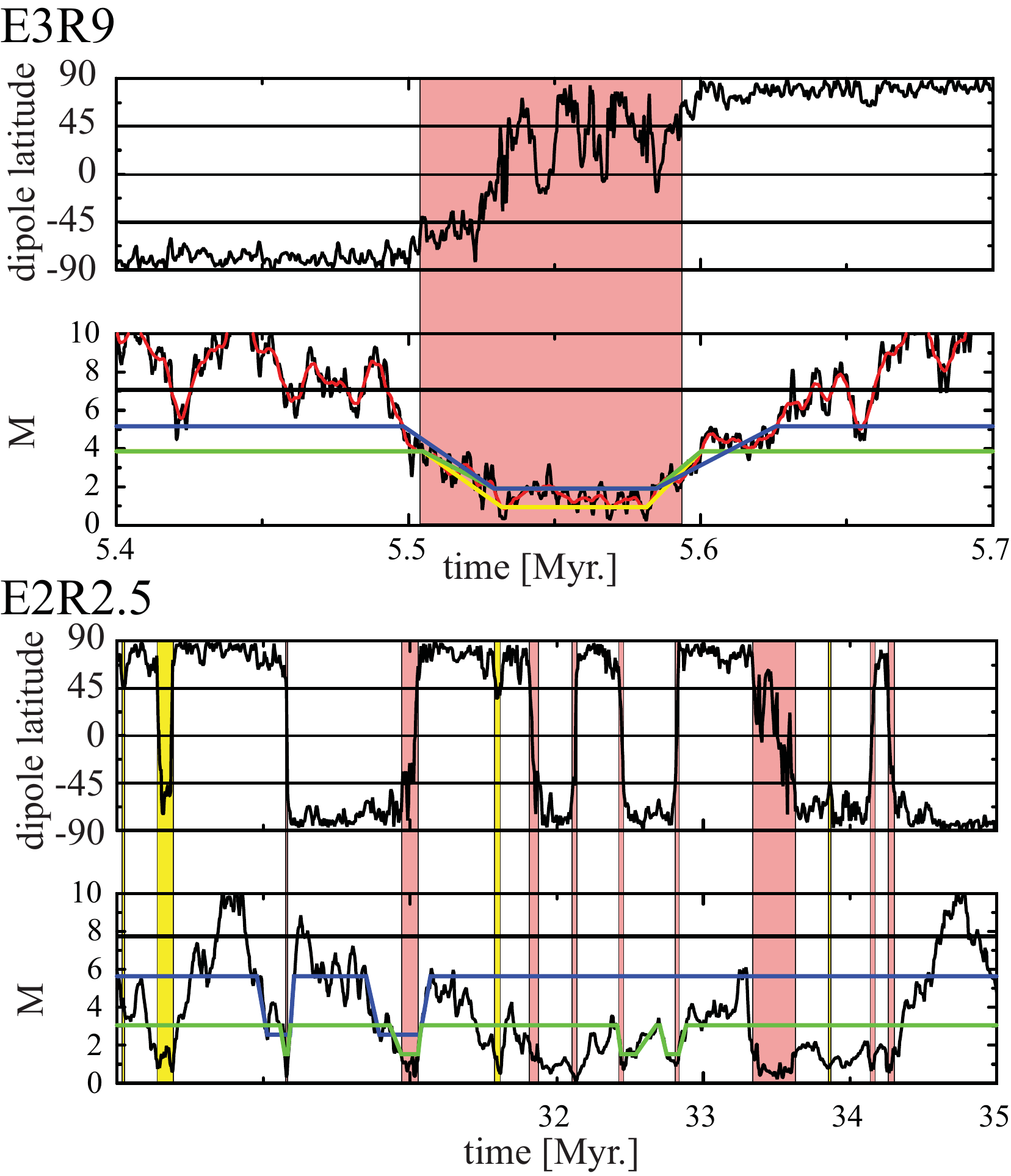}
\caption{Dipole tilt and dipole moment variations during selected
periods in model E3R9 (top) and E2R2.5 (bottom). The red line in the second panel shows
the $5\,$kyr running mean while the other coloured lines illustrate estimates for the decay
and growth periods. Horizontal parts show the threshold dipole moments while the tilted parts
mark the transitions between the assumed high and low dipole moment states.
Model E3R9: blue line for $\MCH=\MH/\MM=0.8$
and green line for $\MCH=0.6$, both using $\MCL=\ML/\MM=0.3$.
The yellow line illustrates the combination ($\MCH=0.60$,$\MCL=0.15$).
Model E2R2.5: blue line for ($\MCH=1.1$,$\MCL=0.5$) and
green line for ($\MCH=0.6$,$\MCL=0.3$).
Yellow and red background stripes once more show the excursions and reversals identified
with the tilt criterion. Thick horizontal black lines in dipole moment panels show the
mean value $\MM_H$ of the high dipole moment state.}
\label{DecGrowProblem}
\end{figure}

\Figref{DecGrowProblem} further illustrates the problem of selecting appropriate thresholds.
When $\MCH$ is too large, short term variations before or after the decisive transition
are regarded as part of the decay or growth period which leads to longer time scale
estimates (compare blue and green line for the growth period in \figref{DecGrowProblem}).
Similar reasoning holds when $\MCL$ is too low
(green and yellow line for the growth period in \figref{DecGrowProblem}).
This suggest that the total decay $\Delta \MC=\MCH-\MCL$ is a key parameter
and the nearly linear dependence on $\MCH$ indicates
that $\dec$ and $\gro$ could be roughly proportional to $\Delta \MC$.
\Figref{NrevE3}e indeed demonstrates that the time scales are
less sensitive to $\MCH$ or $\MCL$ once $\Delta \MC$ is kept fixed.
However, the variation still remains significant and the time to change the
dipole moment by a fixed percentage increases with the dipole moment.
The absolute rather than the relative change is therefore also relevant.
Panel f of figure \figref{NrevE3} shows the dependence of the total mean event durations
$\ovl{\dur}$, normalized with the dipole decay time, on the threshold $\MCH$.
Durations of up to three dipole decay times seem possible which would correspond to roughly
$150\,$kyr for Earth.

\Figref{NrevE3} also illustrates that some properties remain independent of the thresholds:
a) the growth is on average only about $10$\% slower than the decay,
b) both processes are faster than the dipole decay time ($\tau$ smaller than one),
c) grand excursions and reversals have very similar characteristics, and
d) the dipole moment has to decrease to about $30$\% of its mean value
to allow for reversals and grand excursions.

The combination $\MCL=0.30$ and $\MCH=0.60$, somewhat smaller than the values suggested by
the Gaussian model, seems to offer an acceptable compromise for E3R9 that we
adopt for further analysis.
Reasonable mean values for the decay and growth time scales roughly range
from $\ovl{\TC}=0.20$ to $\ovl{\TC}=0.40$ while mean event durations
range from $\ovl{\dur}=1.5$ to $\ovl{\dur}=2.0$.
The analogous analysis for the reversing models E3R23C and E4R106C yields very similar
results. Decay and growth time scales are even more similar than for model E3R9.
Both scales are about $25$\% smaller in model E4R103C
than in models E3R9 and E3R23C.

\begin{figure}
\centering
\includegraphics[draft=false,width=14cm]{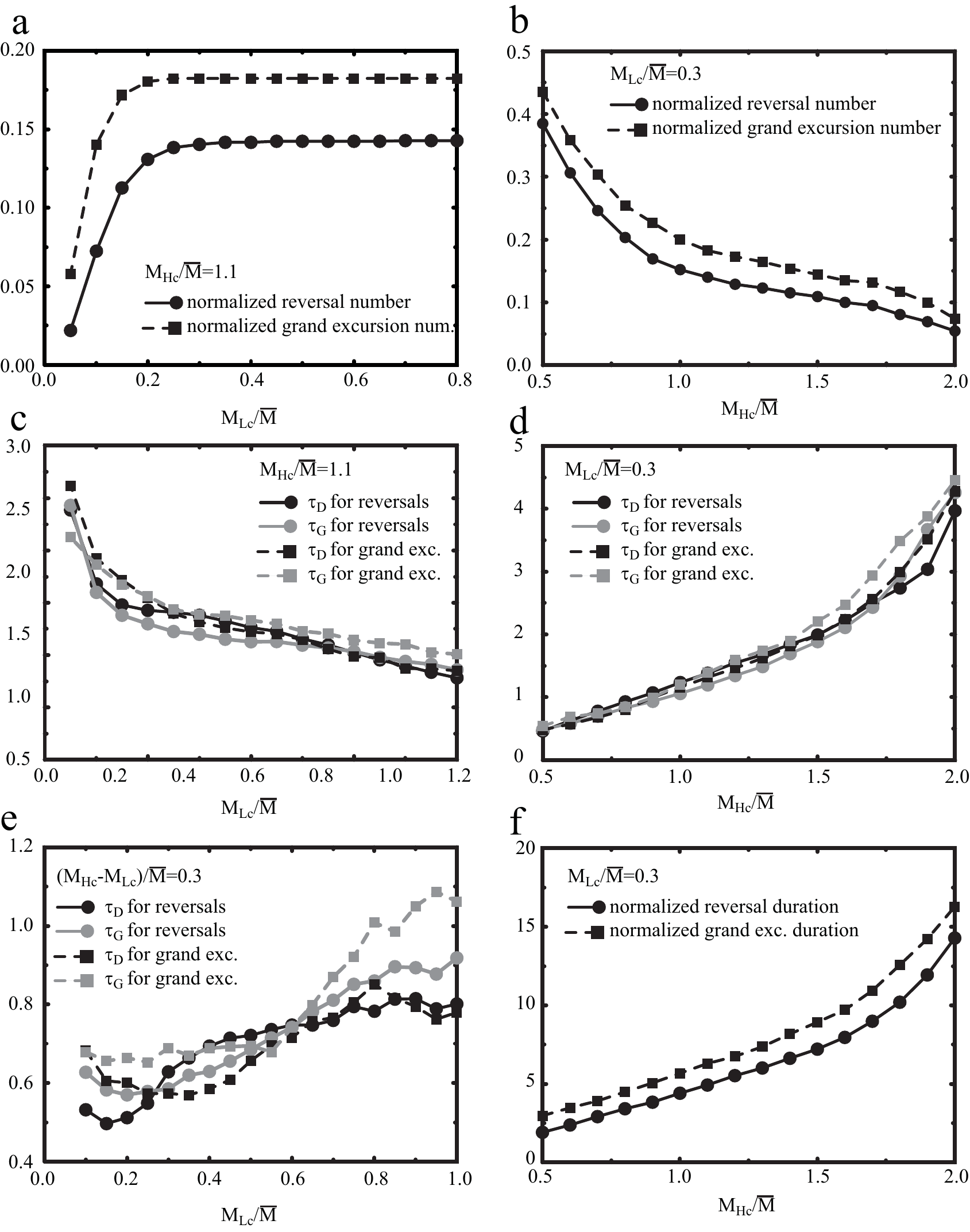}
\caption{Same as \figref{NrevE3} but for model E2R2.5.}
\label{NrevE2}
\end{figure}

Not surprisingly, the different characteristic of the large Ekman number models is also reflected
in the reversal and grand excursion timing.
The analysis of the dipole moment distribution yields a mean high dipole moment
state of $\ovl{M}_H=1.5\MM$ for  E2R2.5 and a standard deviation
of $\sigma(M)=0.4\MM$. This suggest an upper threshold of
$\MCH=1.1$ which, however, significantly reduces the
event number as is demonstrated in \figref{NrevE2}b.
The peak in the axial dipole moment distribution that served for defining $\ovl{M}_H$
has obviously little to do with the majority of reversals and excursions.
The statistics only recovers to some degree when choosing significantly lower values
comparable to those used in the other models.
For $\MCH=0.6$ and $\MCL=0.3$, the values we adopt for further
analysis, about 40\% of the reversals and grand excursions identified by the
tilt criterion enter the time scale analysis.
The lower panels in \figref{DecGrowProblem} further illustrate the difficulties in
choosing appropriate threshold values for E2R2.5. From the 9 reversals
identified with the tilt criterion only one remains for the high
threshold value of $\MCH=1.1$ (blue line) while four are retained for $\MCH=0.6$ (green line).
Many reversals and excursions happen during episodes of low dipole field strength
and relatively unstable tilt angles which complicates the clear timing and
separation of events.

\Figref{NrevE2} demonstrates that the dependence of the time scales
on $\MCH$ and $\MCL$ is nevertheless similar to those found for the other models.
Once more reversals and grand excursions show comparable characteristics´
but take significantly longer than in the
smaller Ekman number cases. $\ovl{\dec}$ and $\ovl{\gro}$
now range between $0.5$ and $2.0$ which seems to suggest
that a stop in dynamo action and the subsequent magnetic field
decay may play a role in the reversal process.
However, since the growth time scale is also very similar, the long time scale
seems to be a more general feature of this dynamo and is possibly a consequence
of the low magnetic Reynolds number.
\Figref{NrevE2}f demonstrates that the event durations are also particularly
long, ranging from $3$ to $15$ dipole decay times.
The longer estimates are characteristic for larger values of $\MCH$
where reversals and grand excursions during weaker field periods are not taken
into account.


\section{Time scale statistics}
\label{Statistics}

\begin{table}
{\scriptsize
\centering
\begin{tabular}{c|*{13}{c}}
Model & $\MCH$&$\MCL$&event&$N_{\theta}$&$N$&$\ovl{\tdec}$&$\ovl{\tgro}$&$\ovl{\tlev}$&$\ovl{\tdur}$&
$\ovl{\twai}$&$\ovl{\left(\frac{\dec}{\gro}\right)}$&
min$\left(\frac{\dec}{\gro}\right)$&max$\left(\frac{\dec}{\gro}\right)$\\
\hline\hline
E3R9   &$0.60$&$0.30$&R& $542$& $513$&$18$&$20$& $82$&$112$&$1840$&$0.88$&$0.05$&$7.2$\\
       &      &      &E& $613$& $590$&$18$&$20$& $67$&$ 99$&$1808$&$0.85$&$0.06$&$12.0$\\
\hline
E3R23C &$0.60$&$0.35$&R& $105$& $100$&$21$&$24$& $69$&$101$& $881$&$0.89$&$0.04$&$8.3$\\
       &      &      &E& $104$& $100$&$23$&$23$& $65$&$101$& $909$&$0.81$&$0.08$&$7.1$\\
\hline
E4R106C&$0.60$&$0.30$&R&  $20$&  $20$&$13$&$13$& $62$& $81$& $740$&$0.89$&$0.26$& $3.6$\\
       &      &      &E&  $29$&  $30$&$14$&$11$& $84$&$104$& $613$&$1.20$&$0.17$&$11.1$\\
\hline
E2R2.5 &$0.60$&$0.30$&R&$2956$& $906$&$35$&$33$& $80$&$134$&$2458$&$1.05$&$0.06$&$12.5$\\
       &      &      &E& $948$& $340$&$33$&$39$&$140$&$197$&$2338$&$0.88$&$0.06$& $8.3$\\
\hline
Earth  &$0.75$&$0.30$&R&  --   &   --  &$33$& $8$&  --  &$49$ & $250$&$4$&--&--\\
\end{tabular}
\caption{Properties of reversals and grand excursion (R or E in column 4)
for selected parameters in the four examined Earth-like reversing models and for Earth.
Column six list the reduced event number when including the dipole moment thresholds in the
event definition.
Columns 7 to 10 show the mean decay and growth time scales
$\tdec$ and $\tgro$, mean time spent in the
LDM state $\ovl{\lev}$, mean event duration $\ovl{\dur}$, and
the mean event waiting time $\ovl{\wai}$, all rescaled to kyr
using the magnetic diffusion time scale.
The last three columns list the mean ratio $\ovl{\dec/\gro}$, its
minimum and its maximum.
Earth values are based on \citet{Valet2005} and \citet{Channell2009} and
a mean reversal rate of four per Myr.}
\label{TabStat}
}
\end{table}

\begin{figure}
\centering
\includegraphics[draft=false,width=14cm]{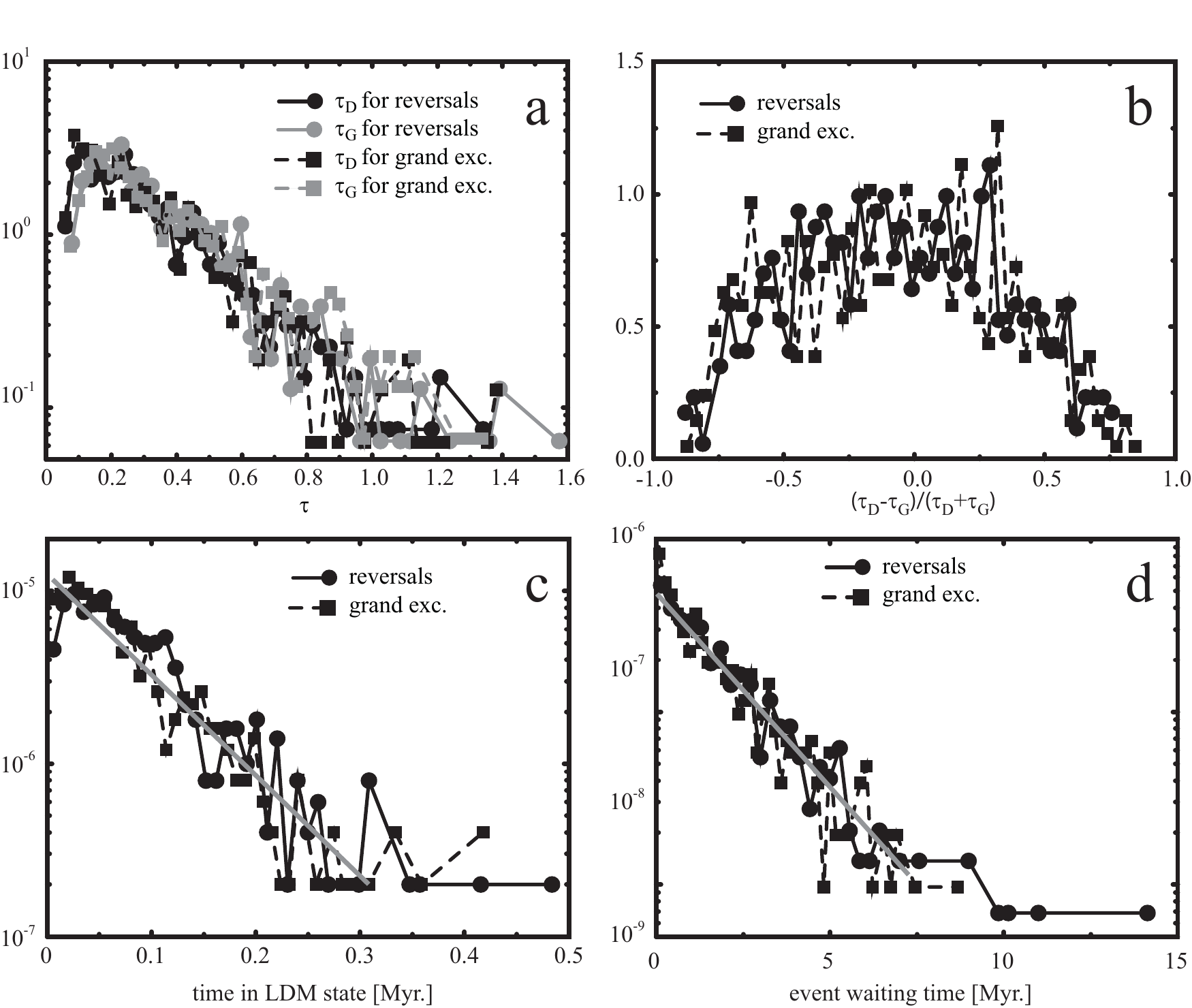}
\caption{Distributions of time scales in model E3R9. Normalized decay and growth time scales
are shown in panel a,
their normalized ratio $R=(\dec-\gro)/(\dec+\gro)$ in panel b,
time $\tlev$ in the LDM state during the event in panel c and event waiting times $\twai$ in panel d.
Tilted grey lines in panel c and d are exponential distributions that are
not the product of a fitting procedure but simple eye-balled suggestions.}
\label{timeE3}
\end{figure}

\begin{figure}
\centering
\includegraphics[draft=false,width=14cm]{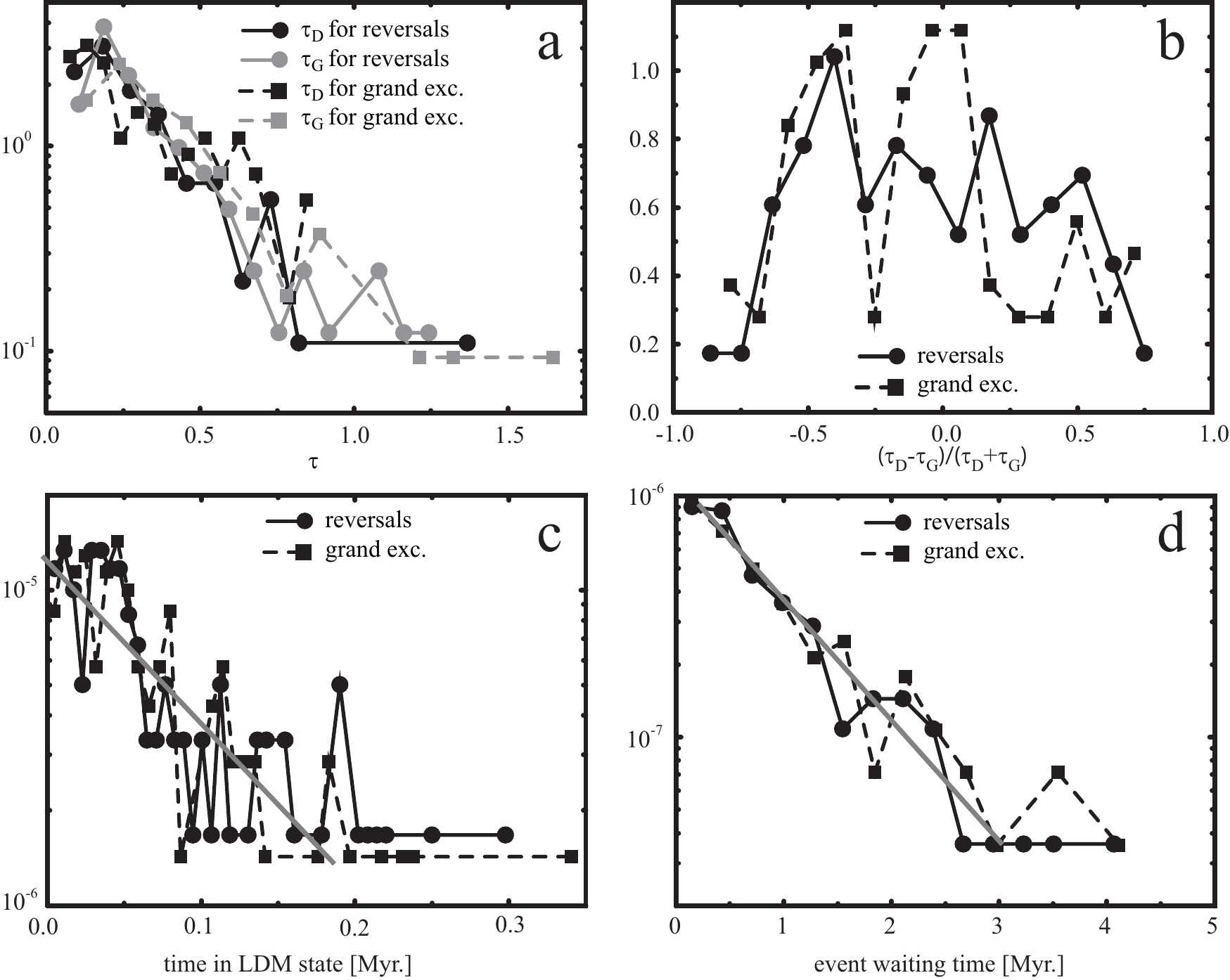}
\caption{Same as \figref{timeE3} but for model E3R23C.}
\label{timeE3C}
\end{figure}

\begin{figure}
\centering
\includegraphics[draft=false,width=14cm]{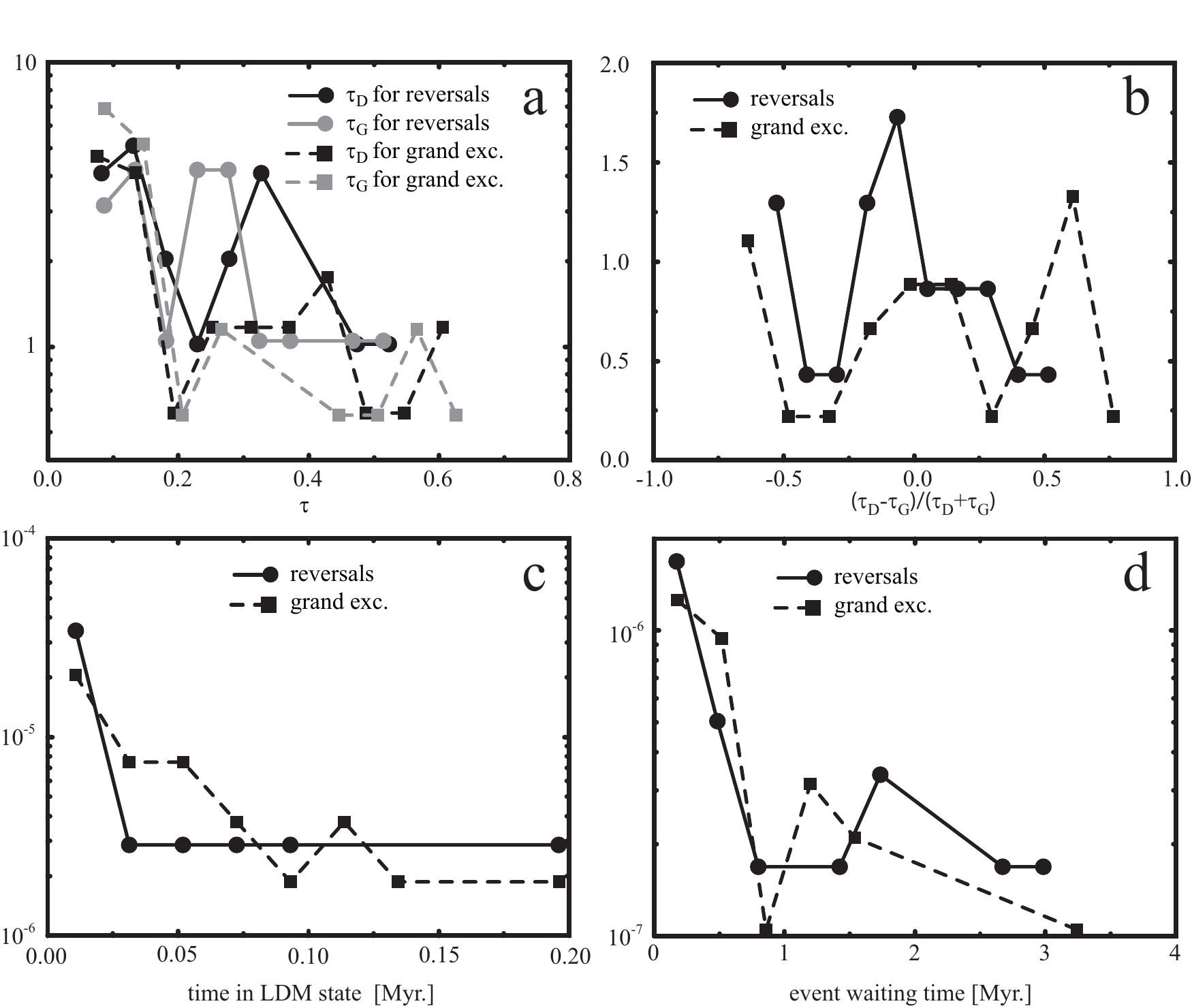}
\caption{Same as \figref{timeE3} but for model E4R106C.}
\label{timeE4C}
\end{figure}

\begin{figure}
\centering
\includegraphics[draft=false,width=14cm]{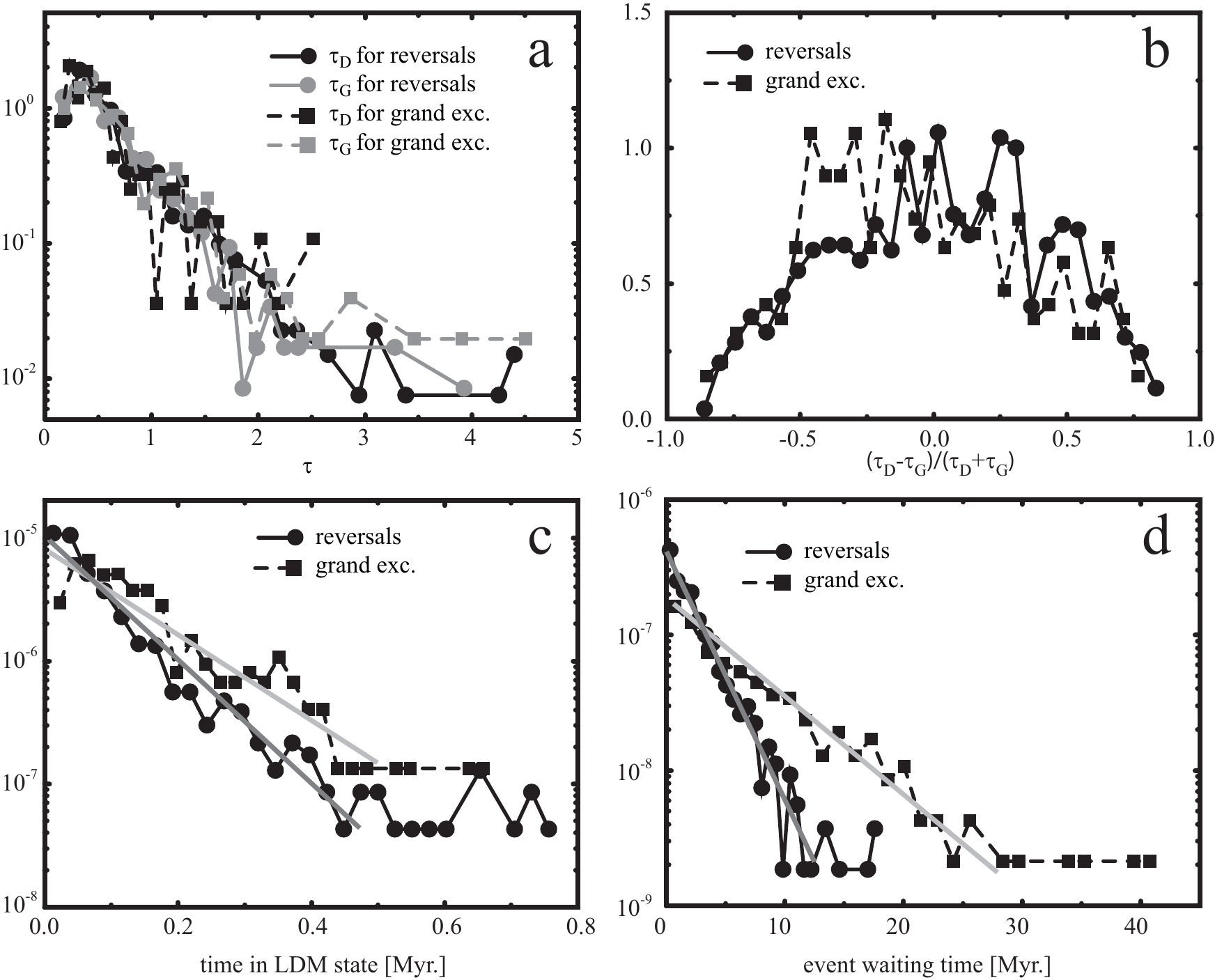}
\caption{Same as \figref{timeE3} but for model E2R2.5.}
\label{timeE2}
\end{figure}

We proceed with more closely analysing the statistics of the different reversal
and excursion time scales for selected threshold values.
\Tabref{TabStat} lists mean time scales and mean, minimum
and maximum ratio of decay to growth time while
figures \ref{timeE3} to \ref{timeE2} show
the probability density distributions.
The event count ranges from $900$ reversals in model E2R2.5 to
$20$ reversals in model E4R106C. The statistics is thus rather poor for the
latter and very good for the former.
Decay and growth
times for reversals and grand excursions are very similar in each of the four models.
The distributions show a lack of very short events simply because the transition
requires some time. The minimum decay and growth time scales are in the order of
a few millenia.
The distributions peak between $0.2$ and $0.4$ dipole decay times and seem to show
an exponential decay
beyond. Naturally, the interpretation is somewhat difficult for E4R106C.
Typical mean decay and growth time scales range
from $11\,$kyr in model E4R106C for $39\,$kyr in model E2R2.5.
Decay time scale estimates for Earth are on the slow side with $33\,$kyr
while the growth is particularly fast with $8\,$kyr.

We have also analysed the ratio of decay to growth time for
all events individually to determine whether the dipole
may decay more slowly than it recovers. A related measure is
$R=(\dec-\gro)/(\dec+\gro)$ which assumes
positive values for a faster recovery and negative
values for a faster decay.
Though decay and growth rate can differ by more than an order of magnitude for
a single event they are very similar on average.
The decay time is typically 10\% faster than the growth time with the
exception of grand excursion in E4R106C and reversals in E2R2.5 (see \tabref{TabStat}).
Panels b of figures \ref{timeE3} to  \ref{timeE2} demonstrate
that the distributions roughly center around $R=0$, are
relatively flat in the middle and decay towards extreme values of both signs.
Earth value of $\tau_D/\tau_G\approx4$ corresponds to $R=0.6$.
When assuming values between $R=0.5$ and $R=0.7$ and an average probability of $0.5$
suggested by \figref{timeE3} to \figref{timeE2} the probability to find one Earth-like
case in the simulations is about $10$\%.
However, the likelihood that five consecutive or very closely spaced reversals
have this property, as suggested by \citet{Valet2005}, is negligibly small.

The total event durations $\ovl{\tdur}$ are up to a factor four longer in the simulations
than suggested for Earth \citep{Valet2005}, mostly because of the significant time
$\ovl{\tlev}$ spend in the LDM state.
$\lev$ distributions
(panels c of figures \ref{timeE3} to \ref{timeE2})
show a less pronounced lack of short durations that the distribution of decay and growth times,
roughly obey an exponential distributions for short and intermediate values,
and suggest a slower decreasing tail of longer events. The
statistical significance of the tail is unclear since is
mainly stays on the one-event level.
Minimum values of $\tlev$ are a millenia or shorter while
the mean ranges from $\ovl{\tlev}=62\,$kyr for reversals in E4R106C to
$\ovl{\tlev}=140\,$kyr for grand excursions in model E2R2.5.
The distributions of the event durations $\tdur$ (not shown) can be understood as
a combination of the distributions of decay, growth, and $\tlev$.
The lack of short durations that is mostly stemms from the time required
for decay and growth lead \citet{Lhuillier2013} to suggest a log-normal distribution.
Mean durations range from $81\,$kyr for reversals in E4R106C to $197\,$kyr for grand excursions
in E2R2.5 (see \tabref{TabStat}).

The event waiting times $\twai$ are most affected by the event
selection process and necessarily increases when dismissing events
not deemed suitable for our time scale analysis.
This effect is rather large for model E2R2.5 but negligible for the other three Earth-like
reversing models.
Mean waiting times range from $0.6\,$Myr for grand excursions in model E4R106C to
$2.5\,$Myr for reversals in model E2R2.5. All distributions seem close to an exponential
decay for shorter to intermediate durations in agreement with
\citet{Wicht2009} and \citet{Lhuillier2013} but also suggest a heavier tail for
very long intervals.

Generally, all reversal and grand excursion distributions are very similar with the
exception of the waiting time distributions for model E2R2.5.
This may partly be attributed to the event selection process but also reflects the
fact that we already count three times more reversals than grand excursions
when using the tilt criterion alone.

\section{Overshoot?}
\label{over}

The magnetic field strength clearly overshoots more normal values at the end of reversals
in the VKS experiment and in some low order dynamical models \citep{Petrelis2009}.
The fast field recovery at the end of a reversals continues to amplitudes
significantly above the mean value and the level of typical fluctuations.
The situation is less clear for paleomagnetic reversals, however.
Only two of the five record discussed by \citet{Valet2005} show an 'overshoot' which
remains within the level of other dipole moment fluctuations.
In a statistical sense an 'overshoot' then simply means that a higher field level is
reached in a particularly short time.

\begin{figure}
\centering
\includegraphics[draft=false,width=14cm]{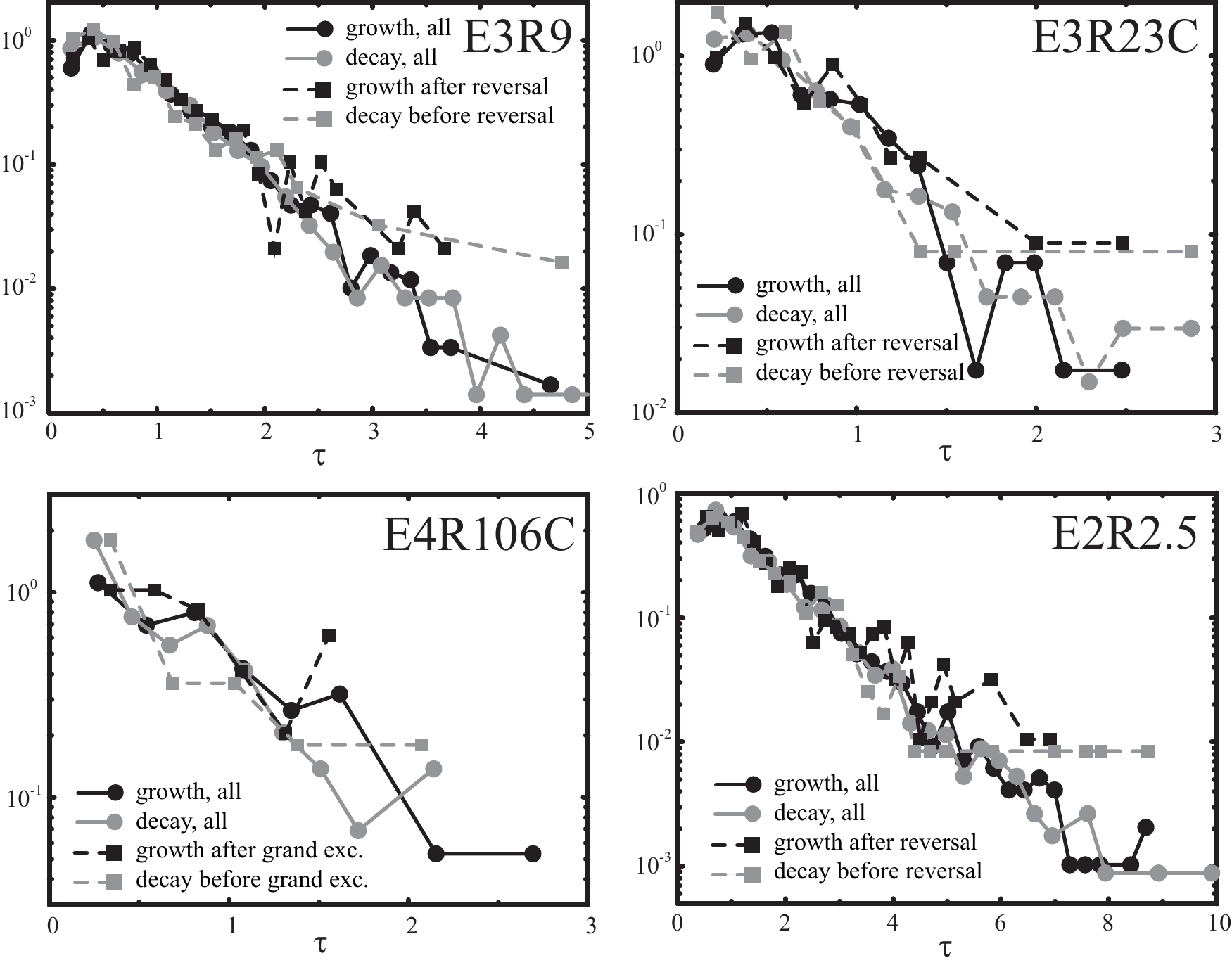}
\caption{Time scale distributions for the four analysed Earth-like
reversing models and thresholds $\MCH=1.6$ and $\MCL=1$.
The individual panels compare all dipole moment variations with those
before and after reversals (models E3R9, E3R23C, E2R2.5) or
grand excursions (mode E4R106C).}
\label{overDist}
\end{figure}

To check whether this is the case in the numerical simulations we
simply raised the thresholds to values within normal dipole moment variations
and show results for the combinations $(\MCL=1.0,\MCH=1.3)$ and
$(\MCL=1.0,\MCH=1.6)$ here.
Once more both the decay and the growth time scales required to
reach the respective other threshold are estimated and we restrict
the analysis not only to reversals (R) and grand excursions (E)
but also more generally consider all (A) dipole moment fluctuations.
The respective mean time scales and time scale
ratios are listed in \tabref{TabOver} where column
four indicates the three event types.
\Figref{overDist} compares the different time scale distributions.

\begin{table}
{\scriptsize
\begin{tabular}{c|*{13}{c}}
Model & $\MCL$&$\MCH$&
event&$N_D$&$N_G$&$\ovl{\tdec}$&$\ovl{\tgro}$&
$\ovl{\left(\frac{\dec}{\gro}\right)}$&min$\left(\frac{\dec}{\gro}\right)$
&max$\left(\frac{\dec}{\gro}\right)$\\
\hline\hline
E3R9   &$1.0$&$1.3$&A&$6275$&$6274$&$26$&$28$&$0.90$&$0.04$&$15.5$\\
       &     &     &R& $339$& $347$&$24$&$29$&$0.82$&$0.06$&$14.9$\\
       &     &     &E& $462$& $485$&$25$&$28$&$0.88$&$0.07$&$11.4$\\
       &     &$1.6$&A&$3205$&$3205$&$45$&$47$&$0.92$&$0.04$&$18.4$\\
       &     &     &R& $325$& $331$&$43$&$51$&$0.82$&$0.05$&$11.0$\\
       &     &     &E& $424$& $438$&$45$&$48$&$0.89$&$0.09$&$16.4$\\
\hline
E3R23C &$1.0$&$1.3$&A& $615$&$616$&$23$&$25$&$0.91$&$0.04$&$14.6$\\
       &     &     &R&  $71$& $73$&$20$&$24$&$0.85$&$0.09$&$10.1$\\
       &     &     &E&  $74$& $77$&$22$&$30$&$0.72$&$0.06$& $4.7$\\
       &     &$1.6$&A& $354$&$355$&$36$&$37$&$0.92$&$0.07$&$15.0$\\
       &     &     &R&  $66$& $69$&$31$&$38$&$0.78$&$0.16$& $7.6$\\
       &     &     &E&  $67$& $71$&$39$&$36$&$0.98$&$0.13$& $7.6$\\
\hline
E4R106C &$1.0$&$1.3$&A& $110$& $110$&$21$&$25$&$0.82$&$0.13$&$14.4$\\
        &     &     &R&   $9$&  $11$&$13$&$25$&$0.76$&$0.16$& $2.2$\\
        &     &     &E&  $16$&  $21$&$16$&$29$&$0.61$&$0.09$& $3.3$\\
        &     &$1.6$&A&  $70$&  $70$&$36$&$44$&$0.84$&$0.12$& $5.7$\\
        &     &     &R&   $9$&  $10$&$38$&$47$&$0.77$&$0.16$& $4.1$\\
        &     &     &E&  $16$&  $20$&$35$&$46$&$0.66$&$0.17$& $6.2$\\
\hline
E2R2.5 &$1.0$&$1.3$&A&$4585$&$4585$&$55$&$58$&$0.94$&$0.07$&$15.3$\\
       &      &    &R& $419$& $443$&$55$&$59$&$0.90$&$0.11$& $7.5$\\
       &      &    &E& $180$& $188$&$47$&$65$&$0.74$&$0.10$& $4.2$\\
       &$1.0$&$1.6$&A&$3445$&$3445$&$79$&$88$&$0.87$&$0.02$&$12.4$\\
       &     &     &R& $411$& $435$&$80$&$92$&$0.85$&$0.06$& $7.8$\\
       &     &     &E& $175$& $185$&$68$&$89$&$0.74$&$0.10$& $8.0$\\
\end{tabular}
\caption{Properties of decay (index D) and growth (index G)
time scales for reversals, grand excursion and all dipole moment
fluctuations (R, E, or A in column four)
for selected parameters in the four examined Earth-like reversing models.
Columns five and six list the event counts, columns seven and
eight the time scales in kyr and columns 9 to 11 the
mean, minimum and maximum ratio of decay and consecutive growth time for
individual events.}
\label{TabOver}
}
\end{table}

The time scales for all three event types are very similar so that the variations
leading in and out of reversals and grand excursions are nothing special.
There is no indication for any overshoot behaviour.
The properties are also similar to the decay and growth from the HDM to the LDM
state discussed in \secref{3Stages}. Time scale values below $\tau=1$ once more
indicate active dynamo processes rather than dipole decay. Distributions lack
short events because of the limited time required to amplify dipole field of normal
or reversed polarity. A pair of decay and following growth time can once more
differ greatly and the decay is on average somewhat faster.
The mean ratio of decay to growth times reaches lower values than
for the HDM-LDM transition. The smallest ratio of $0.61$
for grand excursion in model $E4R106C$, however, is based on rather poor statistics
and a $10$ to $20$\% faster decay seems more typical.

A comparison of time scales for $(\MCH=1.3,\MCL=1)$
in \tabref{TabOver} and $\MCH=0.6,\MCL=0.3)$ in \tabref{TabStat}
confirms that the time required to change the mean dipole moment by
$30$\% increases with the amplitude of the dipole moment
(see also panels e in \figref{NrevE3} and \figref{NrevE2}.

%% file: Discussion.tex
\section{Discussion}
\label{Discussion}

The analysis of dipole moment fluctuations in extremely long
dynamo simulations reveals some interesting
properties of reversals and excursions.
At low Rayleigh numbers, the axial dipole moment distribution
assumes a Gaussian shape with a non zero mean as suggested
by Giant Gaussian models \citep{Constable1988,Hulot1994}.
The mean decreases with growing Rayleigh number until at
some point very low dipole moments become more likely.
The dynamo can then switch from the high dipole moment state
to a new weak field state which is characterized by the fact that
the axial dipole moment assumes a Gaussian distribution with zero mean,
behaves largely like the higher harmonic field contributions,
and frequently switches polarity.

When the dynamo recovers from the weak to the high dipole moment
state, the axial dipole direction is just a matter of chance.
Reversals and grand excursions are therefore equally likely and
have virtually identical properties.
Grand excursions are excursions during which the dipole moment assumes
values typical for the low dipole moment state.
Other excursions still feature large or even inverse tilt angles, in particular
when local magnetic field data are considered (virtual dipole moment) \citep{Wicht2009},
but never a full polarity switch.
The simulations suggest that a drop to $30$\% of the mean field
strength is required to allow for reversals or grand excursions.
This is very similar to the values estimated from paleomagnetic data.

The simulated reversals are thus likely initiated by
particularly large variations in the dipole moment and not
necessarily by any special internal event. The event remain rare since the variations
have to reach the 'outer weak dipole flank' of the Gaussian
distribution.
Similar scenarios are discussed in low-order dynamical
models geared to explain the reversals in the VKS experiment \citep{Petrelis2009},
but there are important differences.
Larger variations that lead beyond the unstable point in the low-order models are
followed by a faster recovery of the dipole moment with opposite polarity and
by an overshoot.
In the numerical simulations, the larger variations leads to the
weak dipole state and since the system tends to linger in this state
for some time it cannot be considered as unstable in the sense of the
low-order models. Consequently, the numerical simulations neither
show a faster recovery not an overshoot.
Both features have been suggested by some paleomagnetic studies
\citep{Valet2005} but are not firmly established.

On average the field decay happens about $10$\% faster than
the growth not only related to reversals and grand excursions
but also for larger amplitude dipole moment variations in general.
Paleomagnetic estimates predict a four times faster growth rate
for reversals and $20$\% faster growth rate for general dipole
moment fluctuation.

Even though there is no significant asymmetry in decay and growth
the time scales for these two phases
are nevertheless similar to those suggested for Earth \citep{Valet2005,Ziegler2011a}.
The overall longer duration of the simulated reversals and excursions is
mostly caused by the pronounced time spent in the lower dipole moment state.
Some paleomagnetic records tend to show a similar lingering at low field intensities
though not to the same extend. The Cobb Mountain event is a prime
example \citep{Channell2009}, the upper Olduvai or even
the Matuyama-Brunhes transitions are other possible candidates
where the field seem to vary around a lower value for some
time \citep{Valet2005,Lund2006,Ziegler2011}.

In terms of a low-dimensional dynamical system reversals
and grand excursions are best described by the three attractors
or states suggested by \citet{Lhuillier2013}, two are the
high dipole moment states of both polarities and the third
is the low dipole moment state that connects both.
The relative stability of these states depends on the system parameters.
For Earth-like reversals, the low dipole state is considerably less
stable than the high dipole state. However, this changes at larger
Rayleigh numbers where the low dipole state seems to
be preferred in the multipolar regime.
Other system parameters like the outer boundary heat flux pattern
\citep{Glatzmaier1999,Kutzner2004,Olson2010,Olson2013}
or the inner thermal boundary conditions \citep{Dharmaraj2012}
are also known to have an impact on the reversal behaviour.
The reversal and grand excursion properties are comparable
for the three smaller Ekman number models explored here.
However, the rate of reversals and grand excursion is about two
times lower for the thermally driven dynamo E3R9 than for the
compositionally driven dynamo E3R23C, despite the fact that
magnetic Reynolds number, local Rossby number and other system
characteristics are very similar.

During the reversals of type 1 described above only the axial dipole moment
decreases while the other field contributions remain largely unaffected.
The larger Ekman number models with $E=2\tp{-2}$ show a second type
of reversals of type 2 where the total field decays for
a some time and then recovers. Common to both types is that reversals
and grand excursions are only possible when the field is low while
other characteristics are rather different.
Type 2 reversals likely play no role for Earth where at least the
multipolar components remain sizable during the events.

The simulations have confirmed that dipole moment variations are
complex and happen on many different time scales.
Reversals and excursions are therefore highly variable and
a statistical approach is required to access their properties.
This became possible due the long runs presented here which
cover up to a thousand reversals.
The statistical analysis then show that there are indeed reversals in
our simulation where the recovery is faster than the decay
and the dipole moment overshoots its mean at the end
of the event, but they are a rare the exception.